\documentclass[12pt]{article}
\usepackage{enumerate}
\usepackage{amsfonts}
\usepackage{amsmath}
\usepackage{amssymb}
\usepackage{amsthm}
\usepackage{latexsym}
\usepackage{color}

\def\H{\mathcal{H}}

\def\S{\mathfrak{S}}
\def\C{\mathfrak{C}}

\def\T{\mathfrak{T}}
\def\B{\mathfrak{B}}

\newcommand{\supp}{\mathrm{supp}}
\newcommand{\rank}{\mathrm{rank}}
\newcommand{\id}{\mathrm{Id}}
\newcommand{\Tr}{\mathrm{Tr}}

\newcommand{\shs}{\hspace{1pt}}
\newcounter{defin}  \newcounter{lemma}  \newcounter{theorem}
\newcounter{property} \newcounter{corol}  \newcounter{remark} \newcounter{example}
\newenvironment{lemma}{\par\refstepcounter{lemma}     \textbf{Lemma \thelemma.} }{\rm\par}
\newenvironment{theorem}{\par\refstepcounter{theorem}     \textbf{Theorem \thetheorem.}\ }{\rm\par}
\newenvironment{property}{\par\refstepcounter{property}     \textbf{Proposition \theproperty.}\ }{\rm\par}
\newenvironment{corollary}{\par\refstepcounter{corol}     \textbf{Corollary \thecorol.} }{\rm\par}
\newenvironment{definition}{\par\refstepcounter{defin}     \textbf{Definition \thedefin.}\ }{\rm\par}
\newenvironment{remark}{\par\refstepcounter{remark}     \textbf{Remark \theremark.}}{\rm\par}
\newenvironment{example}{\par\refstepcounter{example}     \textbf{Example \theexample.}}{\rm\par}

\begin{document}


\title{Convergence criterion for the quantum relative entropy and its use}


\author{M.E. Shirokov\footnote{email:msh@mi.ras.ru}\\Steklov Mathematical Institute, Moscow}
\date{}
\maketitle



\begin{abstract} A criterion and necessary conditions for convergence (local continuity) of the quantum relative entropy are obtained.
Some applications of these results are considered. In particular, the preservation of local continuity of the quantum relative entropy under completely positive linear maps is established.
\end{abstract}


\tableofcontents

\section{Introduction}

The quantum relative entropy is one of the basic quantities in quantum information theory
whose fundamental role in description of information and statistical properties of quantum systems
and channels are well known and quite completely described in the literature (see special
surveys \cite{IRE-1,IRE-2,W}  and the corresponding chapters in \cite{O&P,H-SCI,Wilde}).\smallskip

Mathematically,  the quantum relative entropy $D(\rho\|\shs\sigma)$ is a function of a pair $(\rho,\sigma)$ of quantum
states (or, more generally, positive trace class operators) with a rather complex behavior: it is not continuous and may take infinite
values even in the case of finite dimensional quantum systems. One of the main analytical properties of quantum relative entropy is the (joint) lower semicontinuity, which means that
\begin{equation}\label{D-l-s-c+}
\liminf_{n\to+\infty}D(\rho_n\|\shs\sigma_n)\geq D(\rho_0\|\shs\sigma_0)
\end{equation}
for any sequences  $\,\{\rho_n\}$ and $\{\sigma_n\}$ of quantum states converging, respectively,
to states $\rho_0$ and $\sigma_0$ (with possible value $+\infty$ in one or both sides). Naturally, the question arises about the conditions under which
the limit relation
\begin{equation}\label{D-cont+}
\lim_{n\to+\infty}D(\rho_n\|\shs\sigma_n)=D(\rho_0\|\shs\sigma_0)<+\infty
\end{equation}
holds instead of (\ref{D-l-s-c+}). This question, which is important for applications, has been studied in the literature \cite{W,O&P,L-2,Nau,DTL}. It seems that the first result in this direction
is Lemma 4  in \cite{L-2} that states, in particular, that the limit relation (\ref{D-cont+}) holds for the sequences of states
$$
\rho_n=P_n\rho_0P_n,\quad \sigma_n=P_n\sigma_0P_n,
$$
where $\{P_n\}$ is any nondecreasing sequence of projectors converging to the unit operator in
the strong operator topology, provided that $D(\rho_0\|\shs\sigma_0)<+\infty$.

This research was motivated by the following question.\footnote{I am grateful to G.G.Amosov for pointing my attention to this question.} Assume  that  the limit relation (\ref{D-cont+}) holds for some sequences $\,\{\rho_n\}$ and $\{\sigma_n\}$ of states of a composite quantum system $AB$ converging, respectively, to states $\rho_0$ and $\sigma_0$. What can we say about validity of
(\ref{D-cont+}) for the sequences $\,\{[\rho_n]_A\}$ and $\{[\sigma_n]_A\}$ of marginal states?

The search for an answer to the above question led to necessity to get  necessary and sufficient conditions under which
the limit relation (\ref{D-cont+}) holds. In this article these conditions are obtained and used  to prove several concrete results
concerning existence of the limit relation (\ref{D-cont+}) in some special cases. Most importantly, these conditions allowed us to give a positive answer to the question
stated before by establishing the following property of the (extended) quantum relative entropy:
the validity  of (\ref{D-cont+}) for some sequences $\,\{\rho_n\}$ and $\{\sigma_n\}$ of positive trace class operators converging, respectively,
to operators  $\rho_0$ and $\sigma_0$ implies that
\begin{equation*}
\lim_{n\to+\infty}D(\Phi(\rho_n)\|\shs \Phi(\sigma_n))=D(\Phi(\rho_0)\|\shs \Phi(\sigma_0))<+\infty
\end{equation*}
for arbitrary completely positive linear map $\Phi$. This property can be treated as
preserving local continuity of the quantum relative entropy
under action of completely positive linear maps. In fact, it is a corollary (partial case) of more
general result stating  that (\ref{D-cont+}) implies
\begin{equation*}
\lim_{n\to+\infty}D(\varrho_n\|\shs \varsigma_n)=D(\varrho_0\|\shs \varsigma_0)<+\infty
\end{equation*}
for any sequences  $\{\varrho_n\}$  and $\{\varsigma_n\}$ of positive trace class operators converging, respectively,
to operators  $\varrho_0$ and $\varsigma_0$ such that $\varrho_n=\Phi_n(\rho_n)$ and $\varsigma_n=\Phi_n(\sigma_n)$, where
$\{\Phi_n\}$ is  some sequence of  completely positive linear maps with bounded operator norms (in particular, sequence of quantum operations).


\section{Preliminaries}

\subsection{Basic notations}

Let $\mathcal{H}$ be a separable Hilbert space,
$\mathfrak{B}(\mathcal{H})$ the algebra of all bounded operators on $\mathcal{H}$ with the operator norm $\|\cdot\|$ and $\mathfrak{T}( \mathcal{H})$ the
Banach space of all trace-class
operators on $\mathcal{H}$  with the trace norm $\|\!\cdot\!\|_1$. Let
$\mathfrak{S}(\mathcal{H})$ be  the set of quantum states (positive operators
in $\mathfrak{T}(\mathcal{H})$ with unit trace) \cite{H-SCI,Wilde,BSimon}.

Denote by $I_{\mathcal{H}}$ the unit operator on a Hilbert space
$\mathcal{H}$ and by $\id_{\mathcal{\H}}$ the identity
transformation of the Banach space $\mathfrak{T}(\mathcal{H})$.\smallskip

Trace class operators (not only states) will be denoted by the Greek
letters $\rho$, $\sigma$, $\omega$,... All others linear operators
(in particular, unbounded operators) will be denoted by the Latin
letters $A$, $B$, $H$,... For vectors and operators of rank one on a Hilbert space we will
use the Dirac notation $|\phi\rangle$, $|\chi\rangle\langle\psi|$ (in which the action of an operator $|\chi\rangle\langle\psi|$ on a vector $|\phi\rangle$ gives the vector
$\langle\psi|\phi\rangle|\chi\rangle$ \cite{H-SCI,Wilde}.\smallskip

Speaking about convergence of a sequence $\{\rho_n\}\subset\T(\H)$ to an operator $\rho_0\in\T(\H)$ we will always keep in mind the convergence
w.r.t. to the trace norm.\smallskip

We will use the Mirsky inequality
\begin{equation}\label{Mirsky-ineq+}
  \sum_{i=1}^{+\infty}|\lambda^{\rho}_i-\lambda^{\sigma}_i|\leq \|\rho-\sigma\|_1
\end{equation}
valid for any positive operators $\rho$ and $\sigma$ in $\T(\H)$, where  $\{\lambda^{\rho}_i\}_{i=1}^{+\infty}$
and $\{\lambda^{\sigma}_i\}_{i=1}^{+\infty}$ are  sequence
of eigenvalues of $\rho$ and $\sigma$ arranged in the non-increasing order (taking the multiplicity into account) \cite{Mirsky}. Inequality (\ref{Mirsky-ineq+})
directly follows from the result in \cite{Mirsky} in the case when $\rho$ and $\sigma$ are finite rank operators, its validity in the general case can be  proved via
approximation of the operators $\rho$ and $\sigma$ by the sequences of operators $P_n\rho$ and $Q_n\sigma$, where $P_n$ and $Q_n$ are, respectively,
the spectral projectors of $\rho$ and $\sigma$ corresponding to $n$ maximal eigenvalues.\footnote{I am grateful to A.Winter for the corresponding comment.} \smallskip

Let $H$ be a positive (semi-definite)  operator on a Hilbert space $\mathcal{H}$ (we will always assume that positive operators are self-adjoint). Denote by $\mathcal{D}(H)$ the domain of $H$. For any positive operator $\rho\in\T(\H)$ we will define the quantity $\Tr H\rho$ by the rule
\begin{equation}\label{H-fun}
\Tr H\rho=
\left\{\begin{array}{l}
        \sup_n \Tr P_n H\rho\;\; \textrm{if}\;\;  \supp\rho\subseteq {\rm cl}(\mathcal{D}(H))\\
        +\infty\;\;\textrm{otherwise,}
        \end{array}\right.
\end{equation}
where $P_n$ is the spectral projector of $H$ corresponding to the interval $[0,n]$, ${\rm cl}(\mathcal{D}(H))$ is the closure of $\mathcal{D}(H)$ and $\mathrm{supp}\rho$ is the support of the operator $\rho$ -- the closed subspace spanned by the eigenvectors of $\rho$ corresponding to its positive eigenvalues. \smallskip

Any nonzero operator $\rho$ from the positive cone $\T_+(\H)$ of $\T(\H)$ has the spectral decomposition
$$
\rho=\sum_{i=1}^{+\infty}\lambda_i^{\rho}|\varphi_i\rangle\langle\varphi_i|,
$$
where $\{\varphi_i\}$ is an orthonormal system of eigenvectors of $\rho$ and $\{\lambda_i^{\rho}\}$
the corresponding sequence of eigenvalues which can be always chosen  non-increasing. The above decomposition
is not unique if the operator $\rho$ has multiple eigenvalues. This ambiguity can be eliminated if the eigenvectors
corresponding to the same eigenvalues are chosen according to a certain rule, for example, using the following algorithm
of construction of an ordered basic $\{\psi_i\}$ in any given finite-dimensional subspace $\H_0$ of $\H$:
let $\{\tau_i\}$ be a fixed basic in $\H$ and $\psi_1$ the vector $P_0\tau_j/\|P_0\tau_j\|$, where $P_0$ is the projector onto $\H_0$ and $\tau_j$ is the first vector in the basic  $\{\tau_i\}$ such that
$P_0\tau_i\neq0$, let $\psi_2$ be the vector obtained by the same construction with $\H_0$  replaced by  $\H_0\ominus\{c\psi_1\}$, etc.\smallskip

\begin{remark}\label{spectral-pr} Speaking about the spectral projector of a nonzero operator $\rho$ in $\T_{+}(\H)$ corresponding
to its $m$ maximal eigenvalues we will always keep in mind the projector
$$
\sum_{i=1}^{m}|\varphi_i\rangle\langle\varphi_i|,
$$
where $\{\varphi_i\}$ is an orthonormal system of eigenvectors of $\rho$ constructed by the above described rule.
\end{remark}

The \emph{von Neumann entropy} of a quantum state
$\rho \in \mathfrak{S}(\H)$ is  defined by the formula
$S(\rho)=\Tr\eta(\rho)$, where  $\eta(x)=-x\ln x$ if $x>0$
and $\eta(0)=0$. It is a concave lower semicontinuous function on the set~$\mathfrak{S}(\H)$ taking values in~$[0,+\infty]$ \cite{W,H-SCI,L-2}.
The von Neumann entropy satisfies the inequality
\begin{equation}\label{w-k-ineq}
S(p\rho+(1-p)\sigma)\leq pS(\rho)+(1-p)S(\sigma)+h_2(p)
\end{equation}
valid for any states  $\rho$ and $\sigma$ in $\S(\H)$ and $p\in(0,1)$, where $\,h_2(p)=\eta(p)+\eta(1-p)\,$ is the binary entropy \cite{O&P,Wilde,N&Ch}.\smallskip

We will use the  homogeneous extension of the von Neumann entropy to the positive cone $\T_+(\H)$ defined as
\begin{equation}\label{S-ext}
S(\rho)\doteq(\Tr\rho)S(\rho/\Tr\rho)=\Tr\eta(\rho)-\eta(\Tr\rho)
\end{equation}
for any nonzero operator $\rho$ in $\T_+(\H)$ and equal to $0$ at the zero operator \cite{L-2}.\smallskip

By using concavity of the entropy and inequality (\ref{w-k-ineq}) it is easy to show that
\begin{equation}\label{w-k-ineq+}
S(\rho)+S(\sigma)\leq S(\rho+\sigma)\leq S(\rho)+S(\sigma)+H(\{\Tr\rho,\Tr\sigma\})
\end{equation}
for any $\rho$ and $\sigma$ in $\T_+(\H)$, where $H(\{\Tr\rho,\Tr\sigma\})=\eta(\Tr\rho)+\eta(\Tr\sigma)-\eta(\Tr(\rho+\sigma))$
-- the homogeneous extension of the binary entropy to the positive cone in $\mathbb{R}^2$.
\smallskip

We will use repeatedly the following version of Dini's lemma (cf.\cite{Rudin}).\smallskip

\begin{lemma}\label{Dini} \emph{Let $\{a_n\}_{n>0}\subset\mathbb{R}$ be a sequence converging to $\,a_0\in\mathbb{R}$.
Let $\{a_m^n\}_{n\geq0,m>m_0}$ be a double sequence in $\mathbb{R}$ such that $a_m^n\leq a_{m+1}^n$ for all $n\geq0$ and $m\geq m_0$,
$$
\lim_{m\to+\infty}a^m_n= a_n \quad  \forall n\geq 0 \qquad \textrm{and} \qquad \liminf_{n\to+\infty}a^m_n\geq a^{m}_0\quad  \forall m\geq m_0.
$$
Then}
$$
\lim_{m\to+\infty}\sup_{n\geq0}|a_n-a^m_n|=0.
$$
\end{lemma}

\emph{Proof.} Let $\varepsilon>0$ be arbitrary. The assumptions of the lemma imply existence of $n^1_{\varepsilon}>0$
and $m^1_{\varepsilon}>m_0$ such that
$a_n\leq a_0+\varepsilon$ for all $n\geq n^1_{\varepsilon}$ and $a^m_0\geq a_0-\varepsilon$ for all $m\geq m^1_{\varepsilon}$.
Since $\liminf_{n\to+\infty}a^{m^1_{\varepsilon}}_n\geq a^{m^1_{\varepsilon}}_0$, there is $n^2_{\varepsilon}$
such that $a^{m^1_{\varepsilon}}_n\geq a^{m^1_{\varepsilon}}_0-\varepsilon$ for all $n\geq n^2_{\varepsilon}$. Hence,
$$
a^{m}_n\geq a^{m^1_{\varepsilon}}_n\geq a_n-3\varepsilon\qquad \forall n\geq n_{\varepsilon}\doteq\max\{n^1_{\varepsilon},n^2_{\varepsilon}\},\quad \forall m\geq m^1_{\varepsilon}.
$$
Since $\lim_{m\to+\infty}a^m_n= a_n$ for all $n$, there is $m^2_{\varepsilon}$ such that $a^{m}_n\geq a_n-3\varepsilon$
for all $n< n_\varepsilon$ provided that $m\geq m^2_{\varepsilon}$. Thus $\sup_{n\geq0}|a_n-a^m_n|\leq 3\varepsilon$ for all
$m\geq \max\{m^1_{\varepsilon},m^2_{\varepsilon}\}$. $\Box$

\subsection{Lindblad's extension of the quantum relative entropy}

The \emph{quantum relative entropy} for two states $\rho$ and
$\sigma$ in $\mathfrak{S}(\mathcal{H})$ is defined as
\begin{equation*}
D(\rho\shs\|\shs\sigma)=\sum_i\langle
\varphi_i|\,\rho\ln\rho-\rho\ln\sigma\,|\varphi_i\rangle,
\end{equation*}
where $\{\varphi_i\}$ is the orthonormal basis of
eigenvectors of the state $\rho$ and it is assumed that
$D(\rho\,\|\sigma)=+\infty$ if $\,\mathrm{supp}\rho\shs$ is not
contained in $\shs\mathrm{supp}\shs\sigma$ \cite{Ume,W,L-2}.\smallskip

We will use  Lindblad's extension of quantum relative entropy to any positive
operators $\rho$ and
$\sigma$ in $\mathfrak{T}(\mathcal{H})$ defined as
\begin{equation*}
D(\rho\shs\|\shs\sigma)=\sum_i\langle\varphi_i|\,\rho\ln\rho-\rho\ln\sigma\,|\varphi_i\rangle+\Tr\sigma-\Tr\rho,
\end{equation*}
where $\{\varphi_i\}$ is the orthonormal basis of
eigenvectors of the operator  $\rho$ and it is assumed that $\,D(0\|\shs\sigma)=\Tr\sigma\,$ and
$\,D(\rho\shs\|\sigma)=+\infty\,$ if $\,\mathrm{supp}\rho\shs$ is not
contained in $\shs\mathrm{supp}\shs\sigma$ (in particular, if $\rho\neq0$ and $\sigma=0$)
\cite{L-2}. If the extended von Neumann entropy $S(\rho)$ of $\rho$ (defined in (\ref{S-ext})) is finite
then
\begin{equation}\label{re-exp}
D(\rho\shs\|\shs\sigma)=\Tr\rho(-\ln\sigma)-S(\rho)-\eta(\Tr\rho)+\Tr\sigma-\Tr\rho,
\end{equation}
where $\Tr\rho(-\ln\sigma)$ is defined according to the rule (\ref{H-fun}). \smallskip

The function $(\rho,\sigma)\mapsto D(\rho\shs\|\shs\sigma)$ is nonnegative lower semicontinuous and jointly convex on
$\T_+(\H)\times\T_+(\H)$. We will use the following properties of this function:
\begin{itemize}
  \item for any $\rho,\sigma\in\T_+(\H)$ and $c\geq0$ the following equalities  hold:
  \begin{equation}\label{D-mul}
  D(c\rho\shs\|\shs c\sigma)=cD(\rho\shs\|\shs \sigma),\qquad\qquad\qquad\qquad\quad\;
  \end{equation}
\begin{equation}\label{D-c-id}
D(\rho\shs\|\shs c\sigma)=D(\rho\shs\|\shs\sigma)-\Tr\rho\ln c+(c-1)\Tr\sigma;
\end{equation}

  \item for any $\rho,\sigma$ and $\omega$ in $\T_+(\H)$ the following inequalities hold (with possible values $+\infty$ in one or both sides)
\begin{equation}\label{re-ineq}
D(\rho\shs\|\shs\sigma+\omega)\leq D(\rho\shs\|\shs\sigma)+\Tr\omega, \qquad\qquad\qquad\qquad\qquad\qquad
\end{equation}
\begin{equation}\label{re-2-ineq-conc}
D(\rho+\sigma\|\shs\omega)\geq D(\rho\shs\|\shs\omega)+D(\sigma\shs\|\shs\omega)-\Tr\omega,\qquad\qquad\qquad\quad\;
\end{equation}
\begin{equation}\label{re-2-ineq-conv}
D(\rho+\sigma\|\shs\omega)\leq D(\rho\shs\|\shs\omega)+D(\sigma\shs\|\shs\omega)+ H(\{\Tr\rho,\Tr\sigma\})-\Tr\omega,
\end{equation}
where $H(\{\Tr\rho,\Tr\sigma\})$ is the extended binary  entropy of $\{\Tr\rho,\Tr\sigma\}$ defined in (\ref{w-k-ineq+});
  \item for any $\rho,\sigma,\omega$ and $\vartheta$ in $\T_+(\H)$ the following inequality holds
  \begin{equation}\label{D-sum-g}
    D(\rho+\sigma\shs\|\shs \omega+\vartheta)\leq D(\rho\shs\|\shs \omega)+D(\sigma\shs\|\shs \vartheta),
  \end{equation}
  if $\rho\sigma=\rho\vartheta=\sigma\omega=\omega\vartheta=0$ then
  \begin{equation}\label{D-sum}
    D(\rho+\sigma\shs\|\shs \omega+\vartheta)=D(\rho\shs\|\shs \omega)+D(\sigma\shs\|\shs \vartheta).
  \end{equation}
 \end{itemize}
Inequalities (\ref{re-ineq}), (\ref{re-2-ineq-conc}) and (\ref{re-2-ineq-conv}) are easily proved by using representation (\ref{re-exp}) if the extended von Neumann entropy of the operators $\rho$, $\sigma$  and $\omega$ is finite. Indeed, inequality (\ref{re-ineq}) follows from the operator monotonicity of the logarithm,  inequalities (\ref{re-2-ineq-conc}) and (\ref{re-2-ineq-conv}) follow from the inequalities in (\ref{w-k-ineq+}). In the general case these inequalities can be proved by approximation using Lemma 4 in \cite{L-2}.

Inequality (\ref{D-sum-g}) is a direct corollary of the joint convexity of the relative entropy and identity (\ref{D-mul}). Equality (\ref{D-sum}) follows from the definition \cite{L-2}.

\subsection{Strong convergence of quantum operations}

A \emph{quantum operation} $\Phi$ from a system $A$ to a system
$B$ is a completely positive trace-non-increasing linear map from
$\mathfrak{T}(\mathcal{H}_A)$ into $\mathfrak{T}(\mathcal{H}_B)$.
A trace preserving quantum operation is called  \emph{quantum channel} \cite{H-SCI,Wilde}.   \smallskip

If $\Phi$ is a quantum operation (channel) from $\T(\H_A)$ to $\T(\H_B)$ then the map $\Phi^*:\B(\H_B)\rightarrow\B(\H_A)$ defined by the relation
\begin{equation}\label{dual-r}
\Tr\shs \Phi(\rho)B=\Tr\shs\Phi^*\!(B)\rho\;\quad\forall B\in\B(\H_B)
\end{equation}
is called \emph{dual operation (channel)} to $\Phi$ \cite{H-SCI,B&R,R&S}. \smallskip

In analysis of infinite-dimensional quantum systems the notion of strong convergence of quantum operations is widely used \cite{AQC,CSR}.
A sequence $\{\Phi_n\}$ of quantum operations  from $\T(\H_A)$ to $\T(\H_B)$\emph{ strongly converges} to a quantum operation $\Phi_0$ if
\begin{equation}\label{s-c-def}
\lim_{n\rightarrow+\infty}\mathrm{\Phi}_n(\rho)=\mathrm{\Phi}_0(\rho)\quad \forall \rho\in\mathfrak{S}(\mathcal{H}_A).
\end{equation}
By using the result in \cite{D-A} (which states that the trace-norm convergence in (\ref{s-c-def}) is equivalent to the convergence of the sequence $\{\Phi_n(\rho)\}$ to the operator $\Phi_0(\rho)$ in the weak operator topology provided that $\Tr\mathrm{\Phi}_n(\rho)$ tends to $\Tr\mathrm{\Phi}_0(\rho)$) it is easy to show that  (\ref{s-c-def}) is equivalent to
\begin{equation}\label{s-c-def+}
w\shs\textup{-}\lim_{n\rightarrow+\infty}\mathrm{\Phi}_n^*(B)=\mathrm{\Phi}_0^*(B)\quad \forall B\in\mathfrak{B}(\mathcal{H}_B),
\end{equation}
where $\mathrm{\Phi}_n^*$ is the dual operation to the operation $\mathrm{\Phi}_n$ defined in (\ref{dual-r}) and
$\,w\shs\textup{-}\!\lim$  denotes the limit w.r.t.  the weak operator topology.\footnote{The weak operator topology is essential here. There is another type
of convergence, called strong$*$ convergence in \cite{SSC}, which is characterised by validity of limit relation (\ref{s-c-def+}) in the strong operator topology.
The strong$*$ convergence of quantum channels is the weakest convergence with respect to which the unitary dilation is selective continuous \cite[Theorem 1]{SSC}.}

The strong convergence of infinite-dimensional quantum channels and operations has the obvious advantages over the uniform convergence described in \cite{AQC,W-EBN}.
The characterisations of the strong convergence of quantum channels presented in \cite{CSR}
can be  generalized (with necessary modifications) to the case of operations by using the following\smallskip

\begin{lemma}\label{op-ch} \emph{A sequence $\{\Phi_n\}$ of quantum operations from $\T(\H_A)$ to $\T(\H_B)$ strongly converges to a quantum operation $\Phi_0$
if and only if there is a system $C$ and a sequence $\{\widetilde{\Phi}_n\}$ of quantum channels  from $\T(\H_A)$ to $\T(\H_B\oplus\H_C)$ strongly converging to a quantum channel $\widetilde{\Phi}_0$ such that
\begin{equation}\label{phi-n-rep}
\Phi_n(\rho)=P_B\widetilde{\Phi}_n(\rho)P_B\quad \forall \rho\in\T(\H_A),\; \forall n\geq0,
\end{equation}
where $P_B$ is the projector onto the subspace $\H_B$ of $\mathcal{H}_{B}\oplus\mathcal{H}_{C}$.}
\end{lemma}
\smallskip

\emph{Proof.} If representation (\ref{phi-n-rep}) holds then the validity of (\ref{s-c-def}) follows from the validity of the same relation for the sequence $\{\widetilde{\Phi}_n\}$.

Assume that $\{\Phi_n\}$ is a  sequence of quantum operations from $\T(\H_A)$ to $\T(\H_B)$ strongly converging to a quantum operation $\Phi_0$.
Then representation (\ref{phi-n-rep}) holds with the sequence of quantum channels
$$
\widetilde{\Phi}_n(\rho)=\Phi_n(\rho)\oplus [\Tr(I_A-\Phi_n^*(I_B))\rho\shs]\shs\tau
$$
from $\T(\H_A)$ to $\T(\H_B\oplus\H_C)$, where $\tau$ is a pure state in a one-dimensional Hilbert space $\H_C$.
It follows from (\ref{s-c-def+}) that $\Phi_n^*(I_B)$ tends to $\Phi_0^*(I_B)$ in the weak operator topology. Thus,
the validity of (\ref{s-c-def}) implies the validity of the same relation for the sequence $\{\widetilde{\Phi}_n\}$.
$\Box$\smallskip

By using Lemma \ref{op-ch} it is easy to obtain from Theorem 7 and Corollary 8 in \cite{CSR} the following
characterisations of the strong convergence of quantum operations in terms of the Stinespring and Kraus representations.
\smallskip

\begin{property}\label{s-st-c-ch} \emph{A  sequence $\{\Phi_n\}$ of quantum operations from $\T(\H_A)$ to $\T(\H_B)$ strongly converges to a quantum operation $\Phi_0$ if and only if one of the equivalent conditions holds:}
\begin{itemize}
 \item  \emph{there exist a quantum system $E$ and a sequence $\{V_n\}_{n}$  of contractions  from $\,\mathcal{H}_{A}$ into $\mathcal{H}_{BE}$ strongly converging to a contraction $V_0$  such that $\mathrm{\Phi}_n(\rho)=\mathrm{Tr}_E V_n\rho V^*_n$ for all $\,n\geq0$;}
  \item \emph{there exists a set of sequences $\{A^n_i\}_{n\geq0}$, $i\in I$, of operators from $\mathcal{H}_{A}$ to $\mathcal{H}_{B}$ such that $\mathrm{\Phi}_n(\rho)=\sum_{i\in I} A^n_i\rho [A^n_i]^*$ for all $n\geq0$, $\,s\shs\textup{-}\lim\limits_{n\rightarrow+\infty} A^n_i=A^0_i$ for each $\,i\in I$ and $\,w\shs\textup{-}\lim\limits_{n\rightarrow+\infty} \sum_{i\in I} [A^n_i]^*A^n_i=\sum_{i\in I}[A^0_i]^*A^0_i$.}\footnote{$\,s\shs\textup{-}\!\lim$ and $\,w\shs\textup{-}\!\lim$
      denote the limits in the strong operator and the weak  operator topologies correspondingly.}
\end{itemize}
\end{property}

\section{Convergence criterion for the  relative entropy}

\subsection{General case}

The joint lower semicontinuity of (Lindblad's extension of) the quantum relative entropy means that
\begin{equation}\label{D-l-s-c}
\liminf_{n\to+\infty}D(\rho_n\|\shs\sigma_n)\geq D(\rho_0\|\shs\sigma_0)
\end{equation}
for any sequences  $\,\{\rho_n\}$ and $\{\sigma_n\}$ of operators in $\,\T_+(\H)$ converging, respectively,
to operators  $\rho_0$ and $\sigma_0$. In this section we obtain a necessary and sufficient condition under which
the limit relation
\begin{equation}\label{D-cont++}
\lim_{n\to+\infty}D(\rho_n\|\shs\sigma_n)=D(\rho_0\|\shs\sigma_0)<+\infty
\end{equation}
holds instead of (\ref{D-l-s-c}). We also obtain useful necessary conditions for (\ref{D-cont++}).
\smallskip

The main idea of the convergence criterion described below can be simply explained assuming that $\rho_n$ and $\sigma_n$ are quantum states diagonizable in some orthonormal basis $\{|i\rangle\}$, i.e.
$$
\rho_n=\sum_{i=1}^{+\infty} p_i^n |i\rangle\langle i|\quad \textrm{and} \quad \sigma_n=\sum_{i=1}^{+\infty} q_i^n|i\rangle\langle i|\;\quad \forall n\geq0,
$$
where $\{p_i^n\}_{i=1}^{+\infty}$ and $\{q_i^n\}_{i=1}^{+\infty}$ are probability distributions for each $n\geq0$. In this case the convergence of
the sequences  $\,\{\rho_n\}$ and $\{\sigma_n\}$ to the states $\rho_0$ and $\sigma_0$ means that
\begin{equation}\label{l-conv}
\lim_{n\to+\infty}\sum_{i=1}^{+\infty} |p_i^n-p_i^0|=0\quad \textrm{and}\quad \lim_{n\to+\infty}\sum_{i=1}^{+\infty} |q_i^n-q_i^0|=0,
\end{equation}
while the limit relation
(\ref{D-cont++}) can be written as
\begin{equation}\label{D-cont-cl}
\lim_{n\to+\infty}\sum_{i=1}^{+\infty} a_i^n=\sum_{i=1}^{+\infty} a_i^0<+\infty,
\end{equation}
where $\,a_i^n=p_i^n\ln(p_i^n/q_i^n)+q_i^n-p_i^n\,$ is a nonnegative number for all $i$ and $n$. Assume for simplicity
that the distribution $\{q_i^0\}$ is nondegenerate, i.e.  $q_i^0>0$ for all $i$. Then it is easy to see that the relations in (\ref{l-conv})
imply that
\begin{equation*}
\lim_{n\to+\infty}\sum_{i=1}^{m} a_i^n=\sum_{i=1}^{m} a_i^0<+\infty
\end{equation*}
for any $m\in\mathbb{N}$. Thus, by using Dini's lemma we conclude that (\ref{D-cont-cl}) holds if and only if
\begin{equation*}
\lim_{m\to+\infty}\sup_{n>n_0}\sum_{i>m} a_i^n=0
\end{equation*}
for some $n_0>0$ (if (\ref{D-cont-cl}) holds then $n_0$ is the minimal $n$ such that $\sum_{i=1}^{+\infty}a_i^n$ is finite).
\smallskip

The main difficulty in proving the convergence criterion for the quantum relative
entropy consists in finding a way to realise the above arguments
in the general case when all the states in the sequences  $\,\{\rho_n\}$ and $\{\sigma_n\}$
do not commute to each other.
\smallskip

Slightly modifying\footnote{The modification consists in omitting the condition $\Tr P^n_{m}\sigma_n>0$ (it is necessary, since now we do
not assume that $\sigma_n\neq0$ for all $n\geq0$).} the notation in \cite{DTL} we say that a double sequence $\{P^n_m\}_{n\geq0,m\geq m_0}$ ($m_0\in\mathbb{N}$) of finite rank projectors\footnote{Here and in what follows we assume that the set projectors contains the zero operator.} is \emph{consistent}
with a sequence $\{\sigma_n\}\subset\T_+(\H)$ converging to an operator $\sigma_0$ if
\begin{equation}\label{P-prop}
\sup_{n\geq0}\rank P^n_{m}<+\infty,\quad P^n_{m}\leq P^n_{m+1}, \quad \bigvee_{m\geq m_0}P^n_m\geq Q_n\quad \textrm{and} \quad s\textrm{-}\!\!\lim_{n\to+\infty}P^n_m=P^0_m
\end{equation}
for all $m\geq m_0$ and $n\geq0$, where  $Q_n$ is the projector onto the support of $\sigma_n$ and the limit in the strong operator topology.
\smallskip

The following lemma is easily proved by using Lemma \ref{Dini} in Section 2.1.\smallskip

\begin{lemma}\label{Dini+} \emph{If $\{P^n_m\}_{n\geq0,m\geq m_0}$ is a double sequence of projectors consistent
with a  sequence $\{\sigma_n\}\subset\T_+(\H)$ converging to an operator $\sigma_0$
then $\,\sup_{n\geq0}\Tr(I_{\H}-P^n_{m})\sigma_n$ tends to zero as $\,m\to+\infty$. }
\end{lemma}
\smallskip

We will use the following strengthened version of the above notion.\smallskip

\begin{definition}\label{scs-def} A double sequence $\{P^n_m\}_{n\geq0,m\geq m_0}$ ($m_0\in\mathbb{N}$) of finite rank  projectors is \emph{completely consistent}
with a sequence $\{\sigma_n\}\subset\T_+(\H)$ converging to an  operator $\sigma_0$ if
the conditions in (\ref{P-prop}) hold and
\begin{equation}\label{P-prop+}
P^n_m\sigma_n=\sigma_nP^n_m,\qquad  \rank P^n_m\sigma_n=\rank P^n_m,\qquad n\,\textrm{-}\!\!\lim_{n\to+\infty}P^n_m=P^0_m
\end{equation}
for all $m\geq m_0$ and $n\geq0$, where the limit in the operator norm topology.\smallskip
\end{definition}

For our purposes, the following observation is essential.\smallskip

\begin{lemma}\label{scs-ex} \emph{For any sequence $\{\sigma_n\}\subset\T_+(\H)$ converging to an  operator $\sigma_0$
there exists a double sequence $\{P^n_m\}_{n\geq0,m\geq m_0}$  of finite rank  projectors  completely consistent
with the sequence $\{\sigma_n\}$.}
\end{lemma}
\smallskip

\emph{Proof.} The existence of a double sequence $\{P^n_m\}_{n\geq0,m\geq m_0}$  of projectors completely consistent
with any given  sequence $\{\sigma_n\}\subset\T_+(\H)\setminus\{0\}$ converging to a nonzero operator $\sigma_0$ is
shown in \cite[Lemma 2]{DTL}  (the property $\rank P^n_m\sigma_n=\rank P^n_m$ follows from the construction used in the proof of this lemma).
If we assume that some of the projectors $P^n_m$ may be the zero operators  then the generalization to the case of any converging
sequence $\{\sigma_n\}$ in $\T_+(\H)$ can be done by using the same construction. $\Box$\smallskip

Now we are able to formulate the main results of this section. \smallskip

\begin{theorem}\label{conv-crit} \emph{Let $\,\{\rho_n\}$ and $\{\sigma_n\}$ be sequences of operators in $\,\T_+(\H)$ converging, respectively,
to operators  $\rho_0$ and $\sigma_0$.}\smallskip

A) \emph{The limit relation
\begin{equation}\label{D-cont}
\lim_{n\to+\infty}D(\rho_n\|\shs\sigma_n)=D(\rho_0\|\shs\sigma_0)<+\infty
\end{equation}
holds if and only if there exist
a double sequence $\{P^n_m\}_{n\geq0,m>m_0}$ of finite rank projectors
completely consistent  with the sequence $\{\sigma_n\}$ and $\,n_0\in\mathbb{N}$
such that}
\begin{equation}\label{D-cont-cr}
\lim_{m\to+\infty}\sup_{n\geq n_0} D\!\left(\bar{P}^n_m\rho_n\bar{P}^n_m\,\|\,\bar{P}^n_m\sigma_n\bar{P}^n_m\right)=0,\quad \bar{P}^n_m=I_{\H}-P^n_m.
\end{equation}

B) \emph{If  (\ref{D-cont}) is valid then (\ref{D-cont-cr}) holds with $\,n_0=\min\{n\in\mathbb{N}\,|\,D(\rho_k\|\shs\sigma_k)<+\infty\,\forall k\geq n\}$  for
\textbf{any} double sequence $\{P^n_m\}_{n\geq0,m>m_0}$ of  finite rank  projectors
consistent  with the sequence $\{\sigma_n\}$.}\smallskip

\emph{Claim A remains  valid with condition (\ref{D-cont-cr}) replaced by the formally weaker condition}
\begin{equation}\label{D-cont-cr-w}
\forall\varepsilon>0\quad \exists m_{\varepsilon}>m_0, n_{\varepsilon}>0\,:\, \sup_{n\geq n_\varepsilon }D\!\left(\bar{P}^{n}_{m_{\varepsilon}}\rho_n\bar{P}^{n}_{m_{\varepsilon}}\,\|\,\bar{P}^{n}_{m_{\varepsilon}}\sigma_n
\bar{P}^{n}_{m_{\varepsilon}}\right)<\varepsilon.
\end{equation}
\end{theorem}

\begin{remark}\label{m-r}  Theorem \ref{conv-crit}A contains a criterion of convergence relation (\ref{D-cont})
while Theorem \ref{conv-crit}B  gives a necessary condition for (\ref{D-cont}) which is stronger than the "only if" part of this criterion.
It is the combination of the "if" part of the convergence criterion in Theorem \ref{conv-crit}A and the necessary condition in Theorem \ref{conv-crit}B
allows us to obtain concrete results concerning convergence of the quantum relative entropy (see Sect.~3.3 and 4).
\end{remark}\smallskip

\emph{Proof.} A) The necessity of condition (\ref{D-cont-cr}) for validity of (\ref{D-cont}) follows from part B of the theorem proved below and Lemma \ref{scs-ex}.

To prove the sufficiency assume, at the first step, that the sequence $\{\rho_n\}$ consists of states and
converges to a state $\rho_0$ and that $\sigma_n\neq0$  for all $n\geq0$.

Consider the sequence of functions $f_n(\varrho)\doteq D(\varrho\shs\|\shs\sigma_n)$, $n\geq0$, on $\S(\H)$.
The arguments from the proof of Proposition 2 in \cite{DTL} show the validity of all the conditions of Corollary 1
in \cite{DTL} for the sequence $\{f_n\}$ with $\S_0=\S(\H)$. By using (\ref{D-mul}) we see that the homogeneous
extension of the function $f_n$ to the cone $\T_+(\H)$ has the form
\begin{equation*}
\tilde{f}_n(\varrho)=D(\varrho\shs\|\shs (\Tr\varrho)\sigma_n),\quad \varrho\in \T_+(\H).
\end{equation*}

Let $\{P^n_m\}_{n\geq0,m\geq m_0}$ be a double sequence of finite rank projectors
completely consistent  with the sequence $\{\sigma_n\}$ such that condition (\ref{D-cont-cr}) holds. We may assume that $n_0=1$.
Note that condition (\ref{D-cont-cr}) and the lower semicontinuity of the function $(\varrho,\varsigma)\mapsto D(\varrho\shs\|\shs\varsigma)$
imply that
\begin{equation}\label{D-cont-cr-0}
\lim_{m\to+\infty}D(\bar{P}^0_m\rho_0\bar{P}^0_m\,\|\,\bar{P}^0_m\sigma_0)=0.
\end{equation}
By the first condition in (\ref{P-prop+}) we have $\,\sigma_n=P^n_m\sigma_n+\bar{P}^n_m\sigma_n=P^n_m\sigma_nP^n_m+\bar{P}^n_m\sigma_n\bar{P}^n_m\,$ for all $n\geq0$ and $m\geq m_0$. So, by using  identities (\ref{D-mul}), (\ref{D-c-id}) and (\ref{D-sum}) we obtain
\begin{equation}\label{f-tmp}
\!\!\begin{array}{rl}
\tilde{f}_n(\bar{P}^n_m\rho_n\bar{P}^n_m)\!\!\!&=D(\bar{P}^n_m\rho_n\bar{P}^n_m\shs\|\shs x_m^n\sigma_n)=D(\bar{P}^n_m\rho_n\bar{P}^n_m\shs\|\shs x_m^n\bar{P}^n_m\sigma_n)+x_m^n\Tr P^n_m\sigma_n\\\\
&=D(\bar{P}^n_m\rho_n\bar{P}^n_m\shs\|\shs \bar{P}^n_m\sigma_n)-x_m^n\ln x_m^n-y_m^n+x_m^n\Tr\sigma_n
\end{array}\!\!
\end{equation}
for all $n\geq0$ and $m\geq m_0$, where $x_m^n=\Tr\bar{P}^n_m\rho_n$ and $y_m^n=\Tr\bar{P}^n_m\sigma_n$.  \smallskip

Conditions (\ref{D-cont-cr}), (\ref{D-cont-cr-0}) and the monotonicity of quantum relative entropy under the quantum operation $\,\varrho\mapsto\Tr\varrho\,$ imply
$$
x_m^n\ln(x_m^n/y_m^n)+y_m^n-x_m^n=D(\Tr\bar{P}^n_m\rho_n\|\Tr\bar{P}^n_m\sigma_n)\leq  D(\bar{P}^n_m\rho_n\bar{P}^n_m\shs\|\shs \bar{P}^n_m\sigma_n)\leq\varepsilon_m
$$
for all $n\geq0$ and $m\geq m_0$, where $\{\varepsilon_m\}$ is some sequence tending to zero as $m\to+\infty$.
Lemma \ref{Dini+} and the above inequality show that
\begin{equation}\label{l-n-r}
\lim_{m\to+\infty}\sup_{n\geq0} x_m^n=\lim_{m\to+\infty}\sup_{n\geq0} y_m^n=0.
\end{equation}
Thus, it follows from  (\ref{f-tmp}) that condition (\ref{D-cont-cr}) is equivalent to the condition
\begin{equation*}
\lim_{m\to+\infty}\sup_{n\geq n_0} \tilde{f}_n(\bar{P}^n_m\rho_n\bar{P}^n_m)=0.
\end{equation*}
Hence, by Corollary 1 in \cite{DTL} (its applicability  to the sequence $\{f_n\}$ is mentioned before) to prove
(\ref{D-cont}) it suffices to show that
\begin{equation}\label{D-cont-d}
\lim_{n\to+\infty}\tilde{f}_n(P^n_m\rho_nP^n_m)=\tilde{f}_0(P^0_m\rho_0P^0_m)<+\infty
\end{equation}
for all $m\geq m_0$. By noting that $\,\sigma_n=P^n_m\sigma_n+\bar{P}^n_m\sigma_n=P^n_m\sigma_nP^n_m+\bar{P}^n_m\sigma_n\bar{P}^n_m\,$ and using  identities (\ref{D-mul}),(\ref{D-c-id}) and (\ref{D-sum}) it is easy to see that
$$
\tilde{f}_n(P^n_m\rho_nP^n_m)=D(P^n_m\rho_n P^n_m\shs\|\shs P^n_m\sigma_n)-(1-x_m^n)\ln(1-x_m^n)+y_m^n-x_m^n\Tr\sigma_n
$$
for all $n\geq0$ and $m\geq m_0$. Since $x_m^n$, $y_m^n$ and $\Tr\sigma_n$ tend, respectively, to $x_m^0$, $y_m^0$ and $\Tr\sigma_0$
as $n\to+\infty$, we conclude that (\ref{D-cont-d}) is equivalent to
\begin{equation}\label{D-cont-d+}
\lim_{n\to+\infty}D(P^n_m\rho_n P^n_m\shs\|\shs P^n_m\sigma_n)=D(P^0_m\rho_0 P^0_m\shs\|\shs P^0_m\sigma_0)<+\infty
\end{equation}
Since $P^n_m\rho_n P^n_m$ tends to $P^0_m\rho_0 P^0_m$ as $\,n\to+\infty\,$ and $\,\sup_{n\geq0}\rank P^n_m\rho_n P^n_m< +\infty$  by the  conditions in (\ref{P-prop}), to prove (\ref{D-cont-d+}) it suffices to show that
\begin{equation}\label{D-cont-d++}
\lim_{n\to+\infty}\|P^n_m \ln (P^n_m\sigma_n)-P^0_m \ln (P^0_m\sigma_0)\|=0
\end{equation}
for any given $m\geq m_0$. By the second condition in (\ref{P-prop+}) we have $\rank P^n_m\sigma_n=\rank P^n_m$ for
all $n\geq0$. Hence the sequence  $\{P^n_m\sigma_n+\bar{P}^n_m\}_n$
consists of bounded nondegenerate operators and converges to the nondegenerate operator $P^0_m\sigma_0+\bar{P}^0_m$
in the operator  norm by the last condition in (\ref{P-prop+}). It follows that $P^n_m\sigma_n+\bar{P}^n_m\geq\epsilon I_{\H}$ for
all $n\geq0$ and some $\epsilon>0$. So, by Proposition VIII.20 in \cite{R&S} the sequence  $\{\ln (P^n_m\sigma_n+\bar{P}^n_m)\}_n$
converges to the operator $\ln (P^0_m\sigma_0+\bar{P}^0_m)$ in the operator  norm. This implies (\ref{D-cont-d++}), since
$$
P^n_m \ln (P^n_m\sigma_n)=P^n_m\ln (P^n_m\sigma_n+\bar{P}^n_m)\quad \forall n\geq0.
$$

By using the results of the first step (including the relations in (\ref{l-n-r})) and identities (\ref{D-mul}) and (\ref{D-c-id})
it is easy to establish the "if" part of claim A in the case $\rho_0\neq0$ and $\sigma_0\neq0$.

\smallskip

Assume now that $\rho_0=0$ and $\sigma_0$ is arbitrary. We have to show that condition (\ref{D-cont-cr}) implies that
\begin{equation}\label{D-cont-0}
\lim_{n\to+\infty}D(\rho_n\|\shs\sigma_n)=D(0\|\shs\sigma_0)=\Tr\sigma_0.
\end{equation}

Let $\varepsilon>0$ and $m$ is such that $D(\bar{P}^n_m\rho_n\bar{P}^n_m\|\bar{P}^n_m\sigma_n)\leq\varepsilon$ for all $n\geq n_0$.
By Lemma \ref{re-l} below we have
$$
D(\rho_n\shs\|\shs\sigma_n)\leq D(P^n_m\rho_nP^n_m\|P^n_m\sigma_n)+D(\bar{P}^n_m\rho_n\bar{P}^n_m\|\bar{P}^n_m\sigma_n)+\Tr\rho_n\ln2.
$$
Note that  relation (\ref{D-cont-d++}) is valid in this case by the same arguments (in the case $\sigma_0=0$ we have $P_m^0=0$ for all $m\geq m_0$ and, hence, for any given $m\geq m_0$ there is $n_m$ such that $P^n_m=P^n_m\ln (P^n_m\sigma_n)=P^0_m\ln (P^0_m\sigma_0)=0$
for all $n\geq n_m$, since we assume that $0\ln 0=0$). It implies (\ref{D-cont-d+}) which means  that  $D(P^n_m\rho_nP^n_m\|P^n_m\sigma_n)$ tends to $D(0\|P^0_m\sigma_0)=\Tr P^0_m\sigma_0$ as $n\to+\infty$. Hence the above inequality shows that
$$
\limsup_{n\to+\infty}D(\rho_n\shs\|\shs\sigma_n)\leq \Tr\sigma_0+\varepsilon.
$$
Since $\varepsilon$ is arbitrary, this implies (\ref{D-cont-0}) due to the lower semicontinuity of the relative entropy.\smallskip

B) Assume that (\ref{D-cont}) holds and $\{P^n_m\}_{n\geq0,m>m_0}$ is a double sequence of  projectors
consistent  with the sequence $\{\sigma_n\}$.  We may assume that $D(\rho_n\|\shs\sigma_n)<+\infty$ for all $n$.\smallskip

Since $D(\rho_n\|\shs\sigma_n)<+\infty$, we have $\supp\rho_n\subseteq\supp\sigma_n$. So,
the second and third conditions in (\ref{P-prop}) imply, by Lemma 4 in \cite{L-2}, that $D(P^n_m\rho_n P^n_m\|\shs P^n_m\sigma_n P^n_m)$ monotonously converges to $D(\rho_n\|\shs\sigma_n)$ as $m\to+\infty$ for all $n\geq0$.
Since the sequences $\{P^n_m\rho_n P^n_m\}_n$ and $\{P^n_m\sigma_nP^n_m\}_n$ converge, respectively,  to the operators $P^0_m\rho_0 P^0_m$ and $P^0_m\sigma_0P^0_m$ by the last  condition in (\ref{P-prop}), we have
$$
\liminf_{n\to+\infty}D(P^n_m\rho_n P^n_m\shs\|\shs P^n_m\sigma_nP^n_m)\geq D(P^0_m\rho_0 P^0_m\shs\|\shs P^0_m\sigma_0P^0_m)
$$
due to the lower semicontinuity of the relative entropy. Thus, Lemma \ref{Dini} in Section 2.1
shows that $D(P^n_m\rho_n P^n_m\|\shs P^n_m\sigma_n P^n_m)$ tends to $D(\rho_n\|\shs\sigma_n)$ as $\,m\to+\infty$ uniformly
on $n\geq0$. This implies (\ref{D-cont-cr}) with $n_0=1$, since
$$
D(P^n_m\rho_n P^n_m\shs\|\shs P^n_m\sigma_nP^n_m)+D(\bar{P}^n_m\rho_n\bar{P}^n_m\|\bar{P}^n_m\sigma_n \bar{P}^n_m)\leq D(\rho_n\|\shs\sigma_n)
$$
for all $n\geq0$ and $m\geq m_0$ by Lemma 3 in \cite{L-2}.

To prove the last claim of the theorem assume that (\ref{D-cont-cr-w}) holds. Take $\varepsilon=1$
and choose the corresponding $m_1$ and $n_1$. Since $D(P^{n}_{m}\rho_{n}P^n_m\|P^n_m\sigma_n)$
is finite for all $m$ and $n$ due to the condition $\rank P^n_m\sigma_n=\rank P^n_m$,
Lemma \ref{re-l} below implies that $D(\rho_n\shs\|\shs\sigma_n)$ is finite for all $n\geq n_1$.
Hence the arguments from the proof of claim B and Lemma 4 in \cite{L-2} imply that  $D(\bar{P}^n_m\rho_n\bar{P}^n_m\|\bar{P}^n_m\sigma_n)$ monotonously tends to zero as $m\to+\infty$ for all $n\geq n_1$.
This allows us to show that (\ref{D-cont-cr}) holds with $n_0=n_1$. $\Box$
\smallskip

\begin{lemma}\label{re-l} \emph{Let $\rho$ and $\sigma$ be operators in $\T_+(\H)$ and $P$
a projector in $\B(\H)$ such that $P\sigma=\sigma P$. Then}
\begin{equation}\label{re-l+}
D(\rho\shs\|\shs\sigma)\leq D(P\rho P\shs\|\shs P\sigma)+D(\bar{P}\rho \bar{P}\shs\|\shs\bar{P}\sigma)+\Tr\rho\ln2.
\end{equation}
\end{lemma}

\emph{Proof.} We may assume that the operators $\rho$ and $\sigma$ are nonzero, since otherwise $"="$ holds trivially in (\ref{re-l+}).

It follows from (\ref{D-sum}) that
$$
D\!\left(\textstyle\frac{1}{2}\rho +\textstyle\frac{1}{2}U\rho U^*\shs\|\shs \sigma\right)=D(P\rho P+\bar{P}\rho \bar{P}\shs\|\shs\sigma)=D(P\rho P\shs\|\shs P\sigma)+D(\bar{P}\rho \bar{P}\shs\|\shs\bar{P}\sigma),
$$
where $\bar{P}=I_{\H}-P$ and $U=2P-I_{\H}$ is a unitary operator. By Proposition 5.24 in \cite{O&P}
we have
$$
\begin{array}{rl}
D\!\left(\textstyle\frac{1}{2}\rho+\textstyle\frac{1}{2}U\rho U^*\shs\|\shs \sigma\right)\!\!\!& =rD\!\left(\textstyle\frac{1}{2}(\rho/r)+\textstyle\frac{1}{2}U(\rho/r) U^*\shs\|\shs \sigma/r\right)
\\\\&\geq
\textstyle\frac{1}{2}D(\rho\shs\|\shs \sigma)+\textstyle\frac{1}{2}D(U\rho U^*\shs\|\shs \sigma)-rh_2\!\left(\textstyle\frac{1}{2}\right)=D(\rho\shs\|\shs \sigma)-r\ln2,
\end{array}
$$
$r=\Tr\rho>0$, where we have used (\ref{D-mul}) and taken into account that  $U\sigma U^*=\sigma$ $\Box$. \medskip

\subsection{The function $\rho\mapsto D(\rho\shs\|\shs\sigma)$}

In this subsection we present a criterion and necessary conditions for convergence of the quantum relative
entropy considered as a function of the first argument. To formulate the corresponding corollary
of Theorem \ref{conv-crit} we need the following two
definitions.\smallskip

\begin{definition}\label{cs-def+} A  sequence $\{P_m\}_{m\geq m_0}$  of finite rank  projectors is \emph{consistent}
with a nonzero operator $\sigma$ in $\T_+(\H)$  if
\begin{equation}\label{P-prop-so}
P_{m}\leq P_{m+1}\quad\forall m\geq m_0\quad \textrm{and} \quad \lim_{m\to+\infty}\Tr P_m\sigma=\Tr \sigma.
\end{equation}
\end{definition}
\smallskip
According to this definition, an arbitrary  nondecreasing sequence $\{P_m\}_{m\geq m_0}$ of finite rank projectors strongly converging to the unit operator
is consistent with any operator $\sigma$.
\smallskip

\begin{definition}\label{scs-def+} A  sequence $\{P_m\}_{m\geq m_0}$  of finite rank  projectors is \emph{completely consistent}
with a nonzero operator $\sigma$ in $\T_+(\H)$  if the conditions in (\ref{P-prop-so}) hold,
\begin{equation*}
P_m\sigma=\sigma P_m\quad \textrm{and} \quad\rank P_m\sigma=\rank P_m\quad\;\forall m\geq m_0.
\end{equation*}
\end{definition}
\smallskip

The simplest example of a sequence of finite rank  projectors  completely consistent
with a nonzero operator $\sigma$ is the sequence $\{P_m\}_{m\geq 1}$, where $P_m$ is the spectral projector of $\sigma$
corresponding to its $m$ maximal eigenvalues (taken the multiplicity into account) and it is assumed that $P_m$ is the projector onto the support of
$\sigma$ if $m>\rank\sigma$. \smallskip

Theorem \ref{conv-crit}  applied to the case when the sequence $\{\sigma_n\}$ consists of a single nonzero operator $\sigma$
gives the following
\smallskip

\begin{corollary}\label{conv-crit-c} \emph{Let $\,\{\rho_n\}$ be a sequence of operators in $\,\T_+(\H)$ converging
to an operator  $\rho_0$ and $\sigma$ an arbitrary nonzero operator in $\,\T_+(\H)$.}\smallskip

A) \emph{The limit relation
\begin{equation}\label{D-cont-c}
\lim_{n\to+\infty}D(\rho_n\|\shs\sigma)=D(\rho_0\|\shs\sigma)<+\infty
\end{equation}
holds if and only if there exist
a  sequence $\{P_m\}_{m>m_0}$ of finite rank projectors
completely consistent  with the operator $\sigma$ and $\,n_0\in\mathbb{N}$
such that}
\begin{equation}\label{D-cont-cr-c}
\lim_{m\to+\infty}\sup_{n\geq n_0} D(\bar{P}_m\rho_n\bar{P}_m\,\|\,\bar{P}_m\sigma\bar{P}_m)=0,\quad \bar{P}_m=I_{\H}-P_m.
\end{equation}

B) \emph{If  (\ref{D-cont-c}) is valid then (\ref{D-cont-cr-c}) holds with $\,n_0=\min\{n\in\mathbb{N}\,|\,D(\rho_k\|\shs\sigma)<+\infty\,\forall k\geq n\}$  for
\textbf{any} sequence $\{P_m\}_{m>m_0}$ of  finite rank  projectors
consistent  with the operator $\sigma$.}\smallskip

\emph{Claim A remains  valid with condition (\ref{D-cont-cr-c}) replaced by the formally weaker condition}
\begin{equation*}
\forall\varepsilon>0\quad \exists m_{\varepsilon}>m_0, n_{\varepsilon}>0\,:\, \sup_{n\geq n_\varepsilon }D(\bar{P}_{m_{\varepsilon}}\rho_n\bar{P}_{m_{\varepsilon}}\,\|\,\bar{P}_{m_{\varepsilon}}\sigma
\bar{P}_{m_{\varepsilon}})<\varepsilon.
\end{equation*}
\end{corollary}

\subsection{Simple applications}

Consider examples of using the convergence criteria in Theorem \ref{conv-crit} and Corollary \ref{conv-crit-c}.   \smallskip

\begin{example}\label{e-1} Let $\,\{\rho_n\}$ and $\{\sigma_n\}$ be sequences of operators in $\,\T_+(\H)$ converging, respectively,
to operators  $\rho_0$ and $\sigma_0$ such that $c\rho_n\leq\sigma_n$ for all $n\geq0$ and some $c>0$.

Assume that $\{P^n_m\}_{n\geq0,m>m_0}$ is a double sequence of  projectors
completely consistent  with the sequence $\{\sigma_n\}$ (it exists by Lemma \ref{scs-ex}). Since
$c \bar{P}^n_m\rho_n \bar{P}^n_m\leq \bar{P}^n_m\sigma_n$ for all $n$ and $m$, by applying inequality (\ref{re-2-ineq-conc})
we obtain
$$
D(c\bar{P}^n_m\rho_n \bar{P}^n_m\|\bar{P}^n_m\sigma_n)\leq D(\bar{P}^n_m\sigma_n\|\bar{P}^n_m\sigma_n)+\Tr\bar{P}^n_m\sigma_n=\Tr\bar{P}^n_m\sigma_n.
$$
Using this inequality, identities (\ref{D-mul}) and (\ref{D-c-id}) and Lemma \ref{Dini+} it is easy to show the
validity of condition (\ref{D-cont-cr}) in this case. Thus, relation (\ref{D-cont}) holds  by Theorem \ref{conv-crit}A.
\end{example}\smallskip

\begin{example}\label{e-2} Corollary 3 in \cite[Section 5.1.3]{DTL} states that
\begin{equation}\label{e-2-a}
\lim_{n\to+\infty}D(\rho_n+\sigma_n\|\shs \omega_n+\vartheta_n)=D(\rho_0+\sigma_0\|\shs \omega_0+\vartheta_0)<+\infty
\end{equation}
for any sequences $\,\{\rho_n\}$, $\{\sigma_n\}$, $\{\omega_n\}$ and $\,\{\vartheta_n\}$ of operators in $\,\T_+(\H)$ converging, respectively,
to operators  $\rho_0$, $\sigma_0$, $\omega_0$ and $\,\vartheta_0$ such that
\begin{equation}\label{e-2-c}
\!\!\lim_{n\to+\infty}D(\rho_n\|\shs\omega_n)=D(\rho_0\|\shs\omega_0)<+\infty\quad \textup{and}
\quad\lim_{n\to+\infty}D(\sigma_n\|\shs\vartheta_n)=D(\sigma_0\|\shs\vartheta_0)<+\infty\!
\end{equation}
Theorem \ref{conv-crit} makes this statement a simple corollary
of the joint convexity of the quantum relative entropy. Indeed,
let $\{P^n_m\}_{n\geq0,m>m_0}$ be a double sequence of  projectors
completely consistent  with the sequence $\{\omega_n+\vartheta_n\}$ (such sequence exists by Lemma \ref{scs-ex}).
Since the sequence $\{P^n_m\}_{n\geq0,m>m_0}$ is consistent with the sequences $\{\omega_n\}$ and $\{\vartheta_n\}$,
condition (\ref{e-2-c}) implies, by Theorem \ref{conv-crit}B, that
\begin{equation}\label{e-2-r}
\lim_{m\to+\infty}\sup_{n\geq n_0} D(\bar{P}^n_m\rho_n\bar{P}^n_m\,\|\,\bar{P}^n_m\omega_n\bar{P}^n_m)=\lim_{m\to+\infty}\sup_{n\geq n_0} D(\bar{P}^n_m\sigma_n\bar{P}^n_m\,\|\,\bar{P}^n_m\vartheta_n\bar{P}^n_m)=0
\end{equation}
for some $n_0>0$, where $\bar{P}^n_m=I_{\H}-P^n_m$. By the joint convexity of the quantum relative entropy (in the form of inequality (\ref{D-sum-g}))
we have
\begin{equation*}
D(\bar{P}^n_m(\rho_n+\sigma_n)\bar{P}^n_m\,\|\,\bar{P}^n_m(\omega_n+\vartheta_n)\bar{P}^n_m)\leq D(\bar{P}^n_m\rho_n\bar{P}^n_m\,\|\,\bar{P}^n_m\omega_n\bar{P}^n_m)+
D(\bar{P}^n_m\sigma_n\bar{P}^n_m\,\|\,\bar{P}^n_m\vartheta_n\bar{P}^n_m).
\end{equation*}
Thus, it follows from (\ref{e-2-r}) that
\begin{equation*}
\lim_{m\to+\infty}\sup_{n\geq n_0} D(\bar{P}^n_m(\rho_n+\sigma_n)\bar{P}^n_m\,\|\,\bar{P}^n_m(\omega_n+\vartheta_n)\bar{P}^n_m)=0.
\end{equation*}
By Theorem \ref{conv-crit}A this implies (\ref{e-2-a}).
\end{example}\medskip

Theorem \ref{conv-crit} is used in Section 4 below to generalize the implication (\ref{e-2-c})$\Rightarrow$(\ref{e-2-a})
to the case of countable sums of operators (under a certain condition).
By applying Theorem \ref{conv-crit} one can also obtain simple proofs of Propositions 2 and 3 in \cite{DTL}.
\smallskip

\begin{example}\label{ex-3} Let $H$ be a positive densely defined operator on a Hilbert space $\H$
such that $\Tr e^{-\beta H}<+\infty$ for some $\beta>0$. If we treat $H$ as the Hamiltonian (energy observable) of a quantum system
described by the space $\H$ then the value of $\Tr H\rho$  (defined according to the rule (\ref{H-fun})) is the mean energy of a state $\rho$ in $\S(\H)$.
Condition $\Tr e^{-\beta H}<+\infty$ can be valid only if $H$ is an unbounded operator having  discrete spectrum of finite multiplicity. It means, in Dirac's notation, that
\begin{equation}\label{H-form}
H=\sum_{k=0}^{+\infty} E_k |\tau_k\rangle\langle\tau_k|,
\end{equation}
where
$\left\{\tau_k\right\}_{k=0}^{+\infty}$ is the orthonormal
basis  of eigenvectors of $H$ corresponding to the \emph{nondecreasing} sequence $\left\{E_k\right\}_{k=0}^{+\infty}$ of its eigenvalues
tending to $+\infty$. Representation (\ref{H-form}) means that
$$
H|\varphi\rangle=\sum_{k=0}^{+\infty} E_k\langle\tau_k|\varphi\rangle |\tau_k\rangle\quad \textrm{for any}\;\;  \varphi\in\H\;\; \textrm{such that}\quad \sum_{k=0}^{+\infty} E^2_k|\langle\tau_k|\varphi\rangle|^2<+\infty.
$$
The last condition determines the domain of $H$.
\smallskip

Corollary \ref{conv-crit-c} can be used for analysis of continuity of the function $\rho\mapsto D(\rho\|\shs\gamma_{\beta})$, where
$\gamma_{\beta}=e^{-\beta H}/\Tr e^{-\beta H}$ is the Gibbs state at inverse temperature $\beta$ \cite{W,O&P}.\smallskip

\begin{property}\label{Gibbs}  \emph{Let $\,\{\rho_n\}$ be a sequence of states in $\,\S(\H)$ converging to a
state $\rho_0$. Let $H$ be the operator defined in (\ref{H-form}) such that $\Tr e^{-\beta H}<+\infty$ for some $\beta>0$.}

A) \emph{If there exists a nondecreasing sequence $\{c_k\}_{k=0}^{+\infty}$ of positive numbers tending to $+\infty$ such that
\begin{equation}\label{G-cont-cond}
  \sup_{n\geq n_0} \Tr H_{\{c_k\}}\rho_n<+\infty,\quad \textit{where} \quad H_{\{c_k\}}=\sum_{k=0}^{+\infty} c_kE_k |\tau_k\rangle\langle\tau_k|,
\end{equation}
for some $n_0\geq 0$ then}
\begin{equation}\label{G-cont}
\lim_{n\to+\infty}D(\rho_n\|\shs\gamma_{\beta})=D(\rho_0\|\shs\gamma_{\beta})<+\infty.
\end{equation}

B) \emph{If the operator $H$ satisfies the condition $\Tr e^{-\beta' H}<+\infty$ for any $\beta'>0$, $\sup_{n} \Tr H\rho_n<+\infty$ and the limit relation
(\ref{G-cont}) holds then there exists a nondecreasing sequence $\{c_k\}_{k=0}^{+\infty}$ of positive numbers tending to $+\infty$ such that
(\ref{G-cont-cond}) is valid with $\,n_0=0$.}
\end{property}\smallskip

\emph{Proof.} It is easy to see that the sequence of projectors
\begin{equation*}
P_m=\sum_{k=0}^{m-1} |\tau_k\rangle\langle\tau_k|
\end{equation*}
is completely consistent with the state $\gamma_{\beta}$ (Definition \ref{scs-def+} in Section 3.2).

Since $\Tr e^{-\beta H}<+\infty$, the condition $\sup_{n} \Tr H\rho_n<+\infty$ (which is weaker than the
condition (\ref{G-cont-cond})) implies that $S(\rho_n)$ is finite for all $n\geq0$ by Proposition 1 in \cite{EC}.
Hence $S(\bar{P}_m\rho_n\bar{P}_m)$, where
$\bar{P}_m=I_{\H}-P_m$, is finite for all $m$ and $n\geq0$ by Lemma 3 in \cite{L-2}.
So, it follows from (\ref{re-exp}) that
\begin{equation*}
\begin{array}{c}
\;\;D(\bar{P}_m\rho_n\bar{P}_m\,\|\,\bar{P}_m\gamma_{\beta})= \Tr\bar{P}_m\rho_n\bar{P}_m(-\ln \bar{P}_m\gamma_{\beta})-S(\bar{P}_m\rho_n\bar{P}_m)-\eta(\Tr\bar{P}_m\rho_n)\\\\\displaystyle+\Tr\bar{P}_m(\gamma_{\beta}-\rho_n)
=\beta\sum_{k=m}^{+\infty}E_k\langle\tau_k|\rho_n|\tau_k\rangle-S(\bar{P}_m\rho_n\bar{P}_m)-\eta(\Tr\bar{P}_m\rho_n)\\\\+\,(\ln C_\beta-1)\Tr\bar{P}_m\rho_n + \Tr\bar{P}_m\gamma_{\beta},
\end{array}
\end{equation*}
where $C_\beta=\Tr e^{-\beta H}=\sum_{k=0}^{+\infty}e^{-\beta E_k}$. Hence, since $\,\sup_n\Tr\bar{P}_m\rho_n$ and $\,\Tr\bar{P}_m\gamma_{\beta}$ tend to zero as $m\to+\infty$, condition (\ref{D-cont-cr-c}) in this case  is equivalent to the following one
\begin{equation}\label{G-tmp+}
\limsup_{m\to+\infty}\sup_{n\geq n_0} \left(\beta\sum_{k=m}^{+\infty}E_k\langle\tau_k|\rho_n|\tau_k\rangle-S(\bar{P}_m\rho_n\bar{P}_m)\!\right)\leq0.
\end{equation}

A) If condition (\ref{G-cont-cond}) holds then
\begin{equation*}
\sup_{n\geq n_0} \sum_{k=m}^{+\infty}E_k\langle\tau_k|\rho_n|\tau_k\rangle\leq c_m^{-1}\sup_{n\geq n_0} \sum_{k=m}^{+\infty}c_kE_k\langle\tau_k|\rho_n|\tau_k\rangle\leq c_m^{-1}\sup_{n\geq n_0} \Tr H_{\{c_k\}}\rho_n.
\end{equation*}
Thus, condition (\ref{G-tmp+}) is valid and (\ref{G-cont}) holds by Corollary \ref{conv-crit-c}A.\smallskip

B) Since  $\Tr e^{-\beta' H}<+\infty$ for any $\beta'>0$, the condition  $\sup_{n} \Tr H\rho_n<+\infty$  implies that $S(\rho_n)$ tends to
$S(\rho_0)<+\infty$  as $n\to+\infty$ \cite{W},\cite[Proposition 1]{EC}. By the remark at the begin of the proof  $S(\rho_n)$ is finite for all $n\geq0$. So, Corollary 5B in \cite{DTL} shows that
\begin{equation}\label{G-tmp++}
\lim_{m\to+\infty}\sup_{n\geq0}S(\bar{P}_m\rho_n\bar{P}_m)=0.
\end{equation}
If (\ref{G-cont}) holds then Corollary \ref{conv-crit-c}B implies the validity of (\ref{D-cont-cr-c})
which is equivalent to (\ref{G-tmp+}) in this case. So, by using (\ref{G-tmp++}), we obtain
\begin{equation*}
\lim_{m\to+\infty}\sup_{n\geq n_0}\sum_{k=m}^{+\infty}E_k\langle\tau_k|\rho_n|\tau_k\rangle=0.
\end{equation*}
This guarantees  the existence of a sequence $\{c_k\}_{k=0}^{+\infty}$ with the required properties. $\Box$
\end{example}\smallskip

Proposition \ref{Gibbs}  shows, in particular, that limit relation (\ref{G-cont}) holds provided that
$\sup_{n} \Tr H^\alpha\rho_n<+\infty$ for some $\alpha>1$. Note that the last condition with $\alpha=1$ does not imply (\ref{G-cont}).
Indeed, consider the sequence of states $\rho_n=(1-1/E_n)|\tau_0\rangle\langle\tau_0|+1/E_n|\tau_n\rangle\langle \tau_n|$ converging to the state
$\rho_0=|\tau_0\rangle\langle\tau_0|$. Then $\Tr H\rho_n=E_0(1-1/E_n)+1\leq E_0+1$  and
$$
D(\rho_n\|\shs\gamma_{\beta})=\beta(E_0(1-1/E_n)+1)+\ln C_{\beta}-h_2(1/E_n)\quad \forall n>0,
$$
where $C_{\beta}=\Tr e^{-\beta H}$. We see that $D(\rho_n\|\shs\gamma_{\beta})$ does not tend to $D(\rho_0\|\shs\gamma_{\beta})=\beta E_0+\ln C_{\beta}$.

\section{Preserving convergence for countable sums}

It follows from Corollary 3 in \cite[Section 5.1.3]{DTL} that
\begin{equation*}
\lim_{n\to+\infty}D(\rho^1_n+...+\rho^N_n\|\shs \sigma^1_n+...+\sigma^N_n)=D(\rho^1_0+...+\rho^N_0\|\shs \sigma^1_0+...+\sigma^N_0)<+\infty
\end{equation*}
for any  sequences $\{\rho^1_n\}$,...,$\{\rho^N_n\}$, $\{\sigma^1_n\}$,..., $\{\sigma^N_n\}$ of operators in $\T_+(\H)$ converging, respectively,
to operators  $\rho^1_0$,...,$\rho^N_0$, $\sigma^1_0$,...,$\sigma^N_0$
provided that
\begin{equation}\label{2-rel-c}
\lim_{n\to+\infty}D(\rho^k_n\|\shs\sigma^k_n)=D(\rho^k_0\|\shs\sigma^k_0)<+\infty
\end{equation}
for all $k=1,2,...,N$.  This property can be interpreted as the preservation of convergence of the quantum relative entropy
under (finite) summation.

It is easy to see that the above statement is not valid, in general, in the case $N=+\infty$ even if
the series $\sum_{k}\rho^k_n$  and $\sum_{k}\sigma^k_n$ are well defined trace class operators
converging to the trace class operators $\sum_{k}\rho^k_0$  and $\sum_{k}\sigma^k_0$ correspondingly.

The convergence criterion presented in Theorem \ref{conv-crit} allows us to obtain a sufficient
condition for  validity of the above statement for countable sums.  This condition is a basic
ingredient of the proof of Theorem 2 in the next section.    \smallskip

\begin{property}\label{re-count-sum}  \emph{Let $\,\{\{\rho^k_n\}_{n\geq 0}\}_{k=1}^{+\infty}$ and  $\{\{\sigma^k_n\}_{n\geq 0}\}_{k=1}^{+\infty}$  be  countable sets of converging sequences of operators in $\,\T_+(\H)$ such that
the limit relation (\ref{2-rel-c}) holds,
\begin{equation}\label{K-conv}
  \lim_{n\to+\infty}\sum_{k=1}^{+\infty}\Tr\rho^k_n=\sum_{k=1}^{+\infty}\Tr\rho^k_0<+\infty\quad \textit{and} \quad \lim_{n\to+\infty}\sum_{k=1}^{+\infty}\Tr\sigma^k_n=\sum_{k=1}^{+\infty}\Tr\sigma^k_0<+\infty
\end{equation}
for all $k\in\mathbb{N}$, where $\,\rho^k_0=\lim_{n\to+\infty}\rho^k_n\,$ and $\,\sigma^k_0=\lim_{n\to+\infty}\sigma^k_n$.}\smallskip

\emph{If
\begin{equation}\label{un-sm}
  \lim_{m\to+\infty}\sup_{n\geq n_0}D\!\left.\left(\sum_{k>m}\rho^k_n\,\right\|\shs \sum_{k>m}\sigma^k_n\right)=0
\end{equation}
for some $n_0>0$ then}
\begin{equation}\label{sum-cont}
\lim_{n\to+\infty}D\!\left.\left(\sum_{k=1}^{+\infty}\rho^k_n\,\right\|\shs \sum_{k=1}^{+\infty}\sigma^k_n\right)=D\!\left.\left(\sum_{k=1}^{+\infty}\rho^k_0\,\right\|\shs \sum_{k=1}^{+\infty}\sigma^k_0\right)<+\infty.
\end{equation}\smallskip

\end{property}\smallskip

\begin{remark}\label{re-count-sum-r} If $\,\rho_n^k\rho_n^j=\rho_n^k\sigma_n^j=\sigma_n^k\sigma_n^j=0\,$ for each $n$ and all $k\neq j$ then
by using Dini's lemma one can show that condition (\ref{un-sm}) is necessary for (\ref{sum-cont}), since in this case equality (\ref{D-sum}) implies that
$$
D\!\left.\left(\sum_{k\in K}\rho^k_n\,\right\|\shs \sum_{k\in K}\sigma^k_n\right)=\sum_{k\in K}D(\rho^k_n\|\shs\sigma^k_n)\quad \forall K\subset\mathbb{N},\;\forall n\geq0.
$$
\end{remark}

\emph{Proof.} We may assume w.l.o.g. that $\,\sum_{k=1}^{+\infty}\Tr \rho^k_n<+\infty\,$ and $\,\sum_{k=1}^{+\infty}\Tr \sigma^k_n<+\infty\,$ for all $n\geq 0$.
We may also assume that condition (\ref{un-sm}) holds with $n_0=1$.

For each $n\geq0$ denote the operators $\sum_{k=1}^{+\infty}\rho^k_n$ and $\sum_{k=1}^{+\infty}\sigma^k_n$ by $\varrho_n$ and $\varsigma_n$ correspondingly.
Condition (\ref{K-conv}) guarantees that the sequences $\{\varrho_n\}$ and $\{\varsigma_n\}$ of operators in $\T_+(\H)$  converge, respectively, to the operators $\varrho_0$ and $\varsigma_0$  (this follows from the fact that the convergence of a sequence $\{\varrho_n\}\subset\T_+(\H)$ to an operator $\varrho_0$ in the weak operator topology implies the trace norm convergence provided that $\Tr\varrho_n$ tends to $\Tr\varrho_0$ \cite{D-A}).

Let $\{P^n_m\}_{n\geq0,m>m_0}$ be a double sequence of finite rank projectors
completely consistent  with the sequence $\{\varsigma_n\}$ (such sequence exists by Lemma \ref{scs-ex} in Section 3.1).

For given  arbitrary $\varepsilon>0$ condition (\ref{un-sm}) shows existence of natural $u$ such that
\begin{equation}\label{1-est}
\sup_{n\geq0}D\!\!\left.\left(\sum_{k>u}\rho^k_n\,\right\|\shs \sum_{k>u}\sigma^k_n\right)<\varepsilon.
\end{equation}
For each $n\geq0$ denote the operators $\sum_{k=1}^{u}\rho^k_n$, $\sum_{k>u}\rho^k_n$, $\sum_{k=1}^{u}\sigma^k_n$ and $\sum_{k>u}\sigma^k_n$ by $\varrho^u_n$, $\varrho^v_n$, $\varsigma^u_n$ and $\varsigma^v_n$ correspondingly. By Corollary 3 in \cite{DTL} (which is discussed and reproved in Example \ref{e-2} in Section 3.3) the limit relations in (\ref{2-rel-c}) imply that
\begin{equation*}
\lim_{n\to+\infty}D(\varrho^u_n\|\shs \varsigma^u_n)=D(\varrho^u_0\|\shs \varsigma^u_0)<+\infty.
\end{equation*}
Thus, since the double sequence $\{P^n_m\}_{n\geq0,m>m_0}$ is consistent  with the sequence $\{\varsigma^u_n\}$,
Theorem \ref{conv-crit}B guarantees the existence of $m_{\varepsilon}$ such that
\begin{equation}\label{2-est}
\sup_{n\geq n_\varepsilon} D(\bar{P}^n_{m}\varsigma^u_n\bar{P}^n_m\,\|\shs \bar{P}^n_m\varsigma^u_n\bar{P}^n_m)<\varepsilon
\end{equation}
for all $m\geq m_{\varepsilon}$ and some $n_\varepsilon>0$, where $\bar{P}^n_m=I_{\H}-P^n_m$.

Since  $\,\varrho_n=\varrho^u_n+\varrho^v_n\,$ and $\,\varsigma_n=\varsigma^u_n+\varsigma^v_n$, the joint convexity of the quantum relative entropy (in the form of inequality  (\ref{D-sum-g})) along with (\ref{1-est}) and (\ref{2-est}) implies that
$$
\begin{array}{rl}
D(\bar{P}^n_m\varrho_n\bar{P}^n_m\,\|\shs \bar{P}^n_m\varsigma_n\bar{P}^n_m)\!\!\! & \leq D(\bar{P}^n_m\varrho^u_n\bar{P}^n_m\,\|\shs \bar{P}^n_m\varsigma^u_n\bar{P}^n_m)+ D(\bar{P}^n_m\varrho^v_n\bar{P}^n_m\,\|\shs \bar{P}^n_m\varsigma^v_n\bar{P}^n_m)\\\\ & \leq D(\bar{P}^n_m\varrho^u_n\bar{P}^n_m\,\|\shs \bar{P}^n_m\varsigma^u_n\bar{P}^n_m)+ D(\varrho^v_n\,\|\shs \varsigma^v_n)< 2\varepsilon,
\end{array}
$$
for all $m\geq m_{\varepsilon}$ and $n\geq n_\varepsilon$, where the second inequality follows from Lemma 3 in \cite{L-2}.

Thus, condition (\ref{D-cont-cr-w}) with $\rho_n=\varrho_n$ and $\sigma_n=\varsigma_n$ hold. By  Theorem \ref{conv-crit}A
this implies (\ref{sum-cont}).  $\Box$\smallskip\pagebreak

\section{Preserving convergence under completely positive linear maps}

\subsection{The case of a single map}

A basis property of the quantum relative entropy is its monotonicity under quantum operations (completely positive
trace-non-increasing linear maps), which means that
\begin{equation}\label{m-prop}
D(\Phi(\rho)\|\shs \Phi(\sigma))\leq D(\rho\shs\|\shs\sigma)
\end{equation}
for an arbitrary  quantum operation $\Phi:\T(\H_A)\to\T(\H_B)$ and any operators $\rho$ and $\sigma$ in $\T_+(\H_A)$ \cite{L-REM}.\smallskip

Monotonicity property (\ref{m-prop}) is a necessary ingredient in proving the following theorem that states, in particular, that \emph{local continuity
of the quantum relative entropy is preserved by quantum operations}. \smallskip

\begin{theorem}\label{main}  \emph{Let $\,\{\rho_n\}$ and $\{\sigma_n\}$ be sequences of operators in $\,\T_+(\H_A)$ converging, respectively,
to operators  $\rho_0$ and $\sigma_0$ such that
\begin{equation}\label{in-l-r}
\lim_{n\to+\infty}D(\rho_n\|\shs\sigma_n)=D(\rho_0\|\shs\sigma_0)<+\infty.
\end{equation}
Then
\begin{equation}\label{out-l-r}
\lim_{n\to+\infty}D(\Phi(\rho_n)\|\shs \Phi(\sigma_n))=D(\Phi(\rho_0)\|\shs \Phi(\sigma_0))<+\infty
\end{equation}
for arbitrary completely positive linear map $\Phi:\T(\H_A)\to\T(\H_B)$.}
\end{theorem}\medskip

\emph{Proof.} We may assume that $D(\rho_n\|\shs\sigma_n)<+\infty$ for all $n$.  We may also assume, by relation (\ref{D-mul}), that the map $\Phi$ in trace-non-increasing. It means that
\begin{equation}\label{Kraus}
  \Phi(\rho)=\sum_{k=1}^{+\infty}V_k\rho V_k^*,
\end{equation}
where $\{V_k\}$ is a set of operators from $\H_A$ to $\H_B$ such that $\sum_{k=1}^{+\infty}V_k^*V_k\leq I_{A}$.\smallskip

If $\sigma_0=0$ then (\ref{in-l-r}) can be valid only if $\rho_0=0$ and the limit in (\ref{in-l-r}) is equal to zero. In this case
the validity of (\ref{out-l-r}) trivially follows from the monotonicity property (\ref{m-prop}). So, we will assume  that
$\sigma_0\neq0$.\smallskip

Let $M^*_{\sigma_0}$ be the set of all indexes $i$ such that $\lambda^{\sigma_0}_{i+1}< \lambda^{\sigma_0}_i$, where
$\{\lambda^{\sigma_0}_i\}_{i=1}^{+\infty}$ is the sequence
of eigenvalues of $\sigma_0$ in the non-increasing order (taking the multiplicity into account). Let $m_0$ be the multiplicity
of $\lambda^{\sigma_0}_1$. For each $m\geq m_0$ let $P_m^n$ be the spectral
projector of $\sigma_n$ corresponding to its maximal $\widehat{m}$ eigenvalues\footnote{see Remark \ref{spectral-pr} in Section 2.1 how to avoid the ambiguity of the definition of $P_m^n$ associated with multiple eigenvalues.}, where $\widehat{m}$ is the maximal number in $M^*_{\sigma_0}$ not exceeding  $m$ (if $\rank\sigma_n<\widehat{m}$ then we assume that $P_m^n$ is the projector onto $\supp\sigma_n$).
By using the arguments presented at the end of Section 4.2.1 in \cite{QC}
(based on Theorem VIII.23 in \cite{R&S} and the Mirsky inequality (\ref{Mirsky-ineq+})) it is easy to show that
\begin{equation}\label{P-lim}
   P_m^0=n\,\textup{-}\!\!\lim_{n\to+\infty} P_m^n \qquad \forall m\geq m_0,
\end{equation}
where the limit in the operator norm topology.\smallskip

At the first step we will prove the claim of the theorem assuming that
\begin{equation}\label{p-assump}
  \rank\sigma_n\leq\rank\sigma_0\leq+\infty\quad \forall n.
\end{equation}
If $\,r=\rank\sigma_0<+\infty\,$ then this assumption implies that $P_m^n\sigma_n=\sigma_n$ for all $n$ and $m\geq r$.
\smallskip

By using representation (\ref{Kraus}), Lemma \ref{V-lemma}  below and Proposition \ref{re-count-sum} in Section 4 we see that to prove (\ref{out-l-r}) it suffices to show that
\begin{equation}\label{vir}
 \lim_{j\to+\infty} \sup_{n\geq0} D(\Delta_j(\rho_n)\|\shs \Delta_j(\sigma_n))=0,
\end{equation}
where $\Delta_j(\varrho)=\sum_{k>j}V_k\varrho V_k^*\,$ is a quantum operation for any natural $j$.\smallskip

Let $U_m^n=2P_m^n-I_{\H}$ and $\bar{P}_m^n=I_{\H}-P_m^n$. Then
\begin{equation*}
 P_m^n\rho_n P_m^n+\bar{P}_m^n\rho_n \bar{P}_m^n=\textstyle\frac{1}{2}(\rho_n+U_m^n\rho_n[U_m^n]^*),\quad \forall n,m.
\end{equation*}
Hence, by using inequality (\ref{re-2-ineq-conc}) and identities (\ref{D-mul}) and (\ref{D-c-id}) we obtain
$$
\begin{array}{c}
D(\Delta_j(P_m^n\rho_n P_m^n+\bar{P}_m^n\rho_n \bar{P}_m^n)\|\shs \Delta_j(\sigma_n))\geq
\frac{1}{2}D(\Delta_j(\rho_n)\|\shs 2\Delta_j(\sigma_n))-\Tr\Delta_j(\sigma_n)\\\\
=\frac{1}{2}\left(D(\Delta_j(\rho_n)\|\shs \Delta_j(\sigma_n))-\Tr\Delta_j(\rho_n)\ln2-\Tr\Delta_j(\sigma_n)\right)\quad \forall n,m,j.
\end{array}
$$
Thus, we have
\begin{equation}\label{cr-est-1}
D(\Delta_j(\rho_n)\|\shs \Delta_j(\sigma_n))\leq 2 D(\Delta_j(P_m^n\rho_n P_m^n+\bar{P}_m^n\rho_n \bar{P}_m^n)\|\shs \Delta_j(\sigma_n))+\varepsilon_j,
\end{equation}
where $\varepsilon_j=\sup_{n\geq0}(\Tr\Delta_j(\rho_n)\ln2+\Tr\Delta_j(\sigma_n))$.  Since $\Tr\Delta_j(\rho_n)$ and $\Tr\Delta_j(\sigma_n)$  tend, respectively, to $\Tr\Delta_j(\rho_0)$ and $\Tr\Delta_j(\sigma_0)$ as $n\to+\infty$ for each $j$ and monotonously decrease to zero as $j\to+\infty$ for each $n\geq0$, by using Lemma \ref{Dini} in Section 2.1 we conclude that $\,\varepsilon_j$ tends to zero as $j\to+\infty$. Since $\sigma_n=P_m^n\sigma_n+\bar{P}_m^n\sigma_n=P_m^n\sigma_nP_m^n+\bar{P}_m^n\sigma_n\bar{P}_m^n$, inequality (\ref{D-sum-g}) implies
\begin{equation}\label{cr-est-2}
 \begin{array}{c}
 D(\Delta_j(P_m^n\rho_n P_m^n+\bar{P}_m^n\rho_n \bar{P}_m^n)\|\shs \Delta_j(\sigma_n))\leq D(\Delta_j(P_m^n\rho_n P_m^n)\|\shs \Delta_j(P_m^n\sigma_n))\qquad
\\\\+D(\Delta_j(\bar{P}_m^n\rho_n \bar{P}_m^n)\|\shs \Delta_j(\bar{P}_m^n\sigma_n))\leq \|\Delta^{m,n}_j\|D(\rho_n\|\shs \sigma_n)
+D(\bar{P}_m^n\rho_n \bar{P}_m^n\|\shs \bar{P}_m^n\sigma_n),
\end{array}
\end{equation}
for all $n$, $m$ and $j$, where $\|\Delta^{m,n}_j\|$ is the (operator) norm of the map $\varrho\mapsto \Delta^{m,n}_j(\varrho)\doteq \Delta_j(P_m^n\varrho P_m^n)$.
The last inequality is due to the monotonicity of the relative entropy under the quantum operations $\|\Delta^{m,n}_j\|^{-1}\Delta^{m,n}_j$  and $\Delta_j$ \cite{L-REM}.

Since $\{P_m^n\}_{n\geq0,m\geq m_0}$ is a double sequence consistent with
the sequence $\{\sigma_n\}$ due to the assumption (\ref{p-assump}), the condition (\ref{in-l-r}) and the assumption $D(\rho_n\|\shs\sigma_n)<+\infty$ for all $n$ imply, by Theorem \ref{conv-crit}B, that
\begin{equation*}
\lim_{m\to+\infty} \sup_{n\geq0}D(\bar{P}_m^n\rho_n \bar{P}_m^n\|\shs \bar{P}_m^n\sigma_n)=0.
\end{equation*}
Thus, it follows from (\ref{cr-est-1}) and (\ref{cr-est-2}) that to prove (\ref{vir}) we have to show that
\begin{equation}\label{1-vi-rel}
\lim_{j\to+\infty} \sup_{n\geq0} \|\Delta^{m,n}_j\|=0
\end{equation}
for any given $m\geq m_0$. To prove (\ref{1-vi-rel}) note that
\begin{equation}\label{p-w-c+}
\lim_{j\to+\infty}\|P_m^nT_jP_m^n\|=0\quad \forall n\geq0,m\geq m_0,
\end{equation}
where $T_j=\sum_{k>j}V_k^*V_k$, since $T_j\rightarrow0$  as $\,j\to+\infty$ in the strong operator topology and
$P_m^n$ is a finite rank projector. Note also that
\begin{equation}\label{1-vi-rel+}
\|P_m^nT_jP_m^n\|\leq (\|P_m^nT_j\|+\|T_j P_m^0 \|)\|P_m^n-P_m^0\|+\|P_m^0T_jP_m^0\|.
\end{equation}
It follows from (\ref{P-lim}) that the first term in  (\ref{1-vi-rel+}) can be made arbitrarily small for any given  $m$ and all $j$
by choosing $n$ large enough. Thus, (\ref{p-w-c+}) and  (\ref{1-vi-rel+}) imply (\ref{1-vi-rel}).\smallskip

To complete the proof we have to remove the assumption (\ref{p-assump}).\smallskip

Let $\,r=\rank\sigma_0<+\infty\,$ and $P_r^n$ be the spectral
projector of $\sigma_n$ defined before (\ref{P-lim}). Due to the limit relation (\ref{P-lim}) with $m=r$ and  Lemma \ref{V-lemma} below  it follows from (\ref{in-l-r}) that
\begin{equation}\label{lr-1}
\lim_{n\to+\infty}D(P_r^n\rho_n P_r^n\|\shs P_r^n\sigma_n)=D(P^0_r\rho_0 P^0_r\|\shs P^0_r\sigma_0)=D(\rho_0\|\shs\sigma_0)<+\infty
\end{equation}
and
\begin{equation}\label{lr-2}
\lim_{n\to+\infty}D(\bar{P}_r^n\rho_n \bar{P}_r^n\|\shs \bar{P}_r^n\sigma_n)=D(\bar{P}^0_r\rho_0 \bar{P}^0_r\|\shs \bar{P}^0_r\sigma_0)=D(0\|\shs0)=0.
\end{equation}

By the first step of the proof (under the assumption (\ref{p-assump})) relation (\ref{lr-1}) implies that
\begin{equation*}
\lim_{n\to+\infty}D(\Phi(P_r^n\rho_n P_r^n)\|\shs \Phi(P_r^n\sigma_n))=D(\Phi(\rho_0)\|\shs\Phi(\sigma_0))<+\infty,
\end{equation*}
while relation (\ref{lr-2}) and the monotonicity property (\ref{m-prop}) show that
\begin{equation*}
\lim_{n\to+\infty}D(\Phi(\bar{P}_r^n\rho_n \bar{P}_r^n)\|\shs\Phi(\bar{P}_r^n\sigma_n))=D(0\|\shs0)=0.
\end{equation*}
By Corollary 3 in \cite{DTL} (which is discussed  in Example \ref{e-2} in Section 3.3) these limit relations imply that
$$
\lim_{n\to+\infty}D(\Phi(P_r^n\rho_n P_r^n+\bar{P}_r^n\rho_n \bar{P}_r^n)\|\Phi(\sigma_n))=D(\Phi(\rho_0)\|\shs\Phi(\sigma_0))<+\infty.
$$
Since $\frac{1}{2}\rho_n\leq\frac{1}{2}(\rho_n+U_r^n\rho_n[U_r^n]^*)=P_r^n\rho_n P_r^n+\bar{P}_r^n\rho_n \bar{P}_r^n$, where $U_r^n=2P_r^n-I_{\H}$,
it follows from the last limit relation and Proposition 2 in \cite{DTL} that
\begin{equation}\label{lr-3}
\lim_{n\to+\infty}\!D\left(\textstyle\frac{1}{2}\Phi(\rho_n)\|\shs \Phi(\sigma_n)\right)=\!D\left(\textstyle\frac{1}{2}\Phi(\rho_0)\|\shs\Phi(\sigma_0)\right)<+\infty.
\end{equation}
By using identities  (\ref{D-mul}) and (\ref{D-c-id}) it is easy to show that (\ref{lr-3})  implies (\ref{out-l-r}). $\Box$\medskip

\begin{lemma}\label{V-lemma}  \emph{Let $\,\{\rho_n\}$ and $\{\sigma_n\}$ be sequences of operators in $\,\T_+(\H)$ converging, respectively,
to operators  $\rho_0$ and $\sigma_0$ such that
\begin{equation}\label{V-in-l-r}
\lim_{n\to+\infty}D(\rho_n\|\shs\sigma_n)=D(\rho_0\|\shs\sigma_0)<+\infty.
\end{equation}
Then
\begin{equation}\label{V-out-l-r}
\lim_{n\to+\infty}D(V_n\rho_nV_n^*\|\shs V_n\sigma_nV_n^*)=D(V_0\rho_0V_0^*\|\shs V_0\sigma_0V_0^*)<+\infty
\end{equation}
for any sequence $\{V_n\}$ of bounded operators from the space $\H$ to a Hilbert space $\H'$
such that
\begin{equation}\label{V-cond}
s\shs\textup{-}\lim_{n\rightarrow\infty}V_n=V_0\quad \textit{and}\quad  s\shs\textup{-}\lim_{n\rightarrow\infty}V^*_nV_n=V_0^*V_0,
\end{equation}
where $\,s\shs\textup{-}\!\lim$  denotes the limit in the strong operator topology.}
\end{lemma}\smallskip

\emph{Proof.} Assume first that $V_n=P$ for all $n\geq0$, where $P$ is an (orthogonal) projector in $\B(\H)$. Note that
\begin{equation}\label{b-rep}
P\omega P+\bar{P}\omega \bar{P}=\textstyle\frac{1}{2}(\omega+U\omega U^*),\quad \forall\omega\in\T_+(\H),
\end{equation}
where $\bar{P}=I_{\H}-P$ and $U=2P-I_{\H}$. Since $U$ is a unitary operator, it follows from (\ref{V-in-l-r}) that
\begin{equation}\label{U-l-r}
\lim_{n\to+\infty}D(U\rho_nU^*\|\shs U\sigma_nU^*)=D(U\rho_0U^*\|\shs U\sigma_0U^*)<+\infty.
\end{equation}
By Corollary 3 in \cite{DTL} relations (\ref{V-in-l-r}) and (\ref{U-l-r}) along with (\ref{D-mul}) and (\ref{b-rep}) imply  that
\begin{equation*}
\lim_{n\to+\infty}D(P\rho_nP+\bar{P}\rho_n\bar{P}\shs\|\shs P\sigma_nP+\bar{P}\sigma_n\bar{P})=D(P\rho_0P+\bar{P}\rho_0\bar{P}\shs\|\shs P\sigma_0P+\bar{P}\sigma_0\bar{P})<+\infty.
\end{equation*}
Since $\,D(P\rho_nP+\bar{P}\rho_n\bar{P}\shs\|\shs P\sigma_nP+\bar{P}\sigma_n\bar{P})=D(P\rho_n P\shs\|\shs P\sigma_n P)+D(\bar{P}\rho_n\bar{P}\shs\|\shs \bar{P}\sigma_n\bar{P})\,$
for all $n\geq0$ (due to equality (\ref{D-sum})), by using this limit relation and the lower semicontinuity of the relative entropy it is easy to show that (\ref{V-out-l-r}) holds with $V_n=P$.

Assume now that $\{V_n\}$ is a sequence of contractions from $\H$ to $\H'$ (the general case is reduced to this one by using equality (\ref{D-mul}), since
condition (\ref{V-cond}) implies that $\sup_n\|V_n\|<+\infty$ by the uniform boundedness principle \cite[Theorem III.9]{R&S}).
Consider the sequence $\{\widehat{V}_n\}$ of isometries from $\H$ to $\H'\oplus\H$ defined by the settings
$$
\widehat{V}_n|\varphi\rangle=V_n|\varphi\rangle\oplus\sqrt{I_{\H}-V_n^*V_n}|\varphi\rangle, \quad \varphi\in\H.
$$
Condition (\ref{V-cond}) implies the
convergence of the sequence $\{\widehat{V}_n\}$ to the isometry $\widehat{V}_0$ in the strong operator topology. Hence,
the operators $\widehat{V}_n\rho_n\widehat{V}_n^*$ and $\widehat{V}_n\sigma_n\widehat{V}_n^*$ tend, respectively, to the
 operators $\widehat{V}_0\rho_0\widehat{V}_0^*$ and $\widehat{V}_0\sigma_0\widehat{V}_0^*$ as $n\to+\infty$ in the trace norm.

It follows from (\ref{V-in-l-r}) that (\ref{V-out-l-r}) with $V_n=\widehat{V}_n$ holds. Since
$V_n\omega V_n^*=P\widehat{V}_n\omega \widehat{V}_n^*P$ for any $\omega$ in $\T_+(\H)$,
where $P$ is the projector onto the subspace $\H'$ of $\H'\oplus\H$, the validity of (\ref{V-out-l-r})
follows from the validity of (\ref{V-out-l-r}) with $V_n=\widehat{V}_n$ by the first part of the proof. $\Box$
\pagebreak

\subsection{The case of a sequence of completely positive linear maps}

The statement of Theorem \ref{main}  can be significantly strengthened.
\smallskip

\begin{theorem}\label{main++} \emph{Let $\,\{\rho_n\}$ and  $\{\sigma_n\}$ be sequences of operators in $\,\T_+(\H_A)$ converging, respectively, to operators  $\rho_0$ and $\sigma_0$ such that}
\begin{equation}\label{in+}
\lim_{n\to+\infty}D(\rho_n\|\shs\sigma_n)=D(\rho_0\|\shs\sigma_0)<+\infty.
\end{equation}

\emph{If  $\,\{\varrho_n\}$ and $\{\varsigma_n\}$ are sequences of operators in $\,\T_+(\H_B)$ converging, respectively, to operators  $\varrho_0$ and $\varsigma_0$
such that $\varrho_n=\Phi_n(\rho_n)$ and $\varsigma_n=\Phi_n(\sigma_n)$ for each $n\neq0$, where $\Phi_n$ is a quantum operation from $A$ to $B$,  then}
\begin{equation}\label{out+}
\lim_{n\to+\infty}D(\varrho_n\|\shs\varsigma_n)=D(\varrho_0\|\shs\varsigma_0)<+\infty.
\end{equation}
\end{theorem}\smallskip

\textbf{Note:} It is essential that we do not assume any properties of the sequence $\{\Phi_n\}$. \smallskip

By  identity (\ref{D-mul}) we may assume in the condition of Theorem \ref{main++} that
$\{\Phi_n\}$  is a sequence of completely positive linear maps from $\T(\H_A)$ to $\T(\H_B)$ with bounded operator norms.\medskip

Theorem \ref{main++}  implies the following result, in which the notion of strong convergence
of sequences of quantum operations described in Section 2.3 is used.\smallskip

\begin{corollary}\label{main++c}  \emph{Let $\,\{\rho_n\}$ and $\{\sigma_n\}$ be sequences of operators in $\,\T_+(\H_A)$ converging, respectively,
to operators  $\rho_0$ and $\sigma_0$ such that (\ref{in+}) holds. Then
\begin{equation}\label{out-l-r-c}
\lim_{n\to+\infty}D(\Phi_n(\rho_n)\|\shs \Phi_n(\sigma_n))=D(\Phi_0(\rho_0)\|\shs \Phi_0(\sigma_0))<+\infty
\end{equation}
for arbitrary sequence $\{\Phi_n\}$ of quantum operations from $A$ to $B$ strongly converging to a quantum operation $\Phi_0$.}
\end{corollary}\smallskip

To obtain Corollary \ref{main++c} from Theorem \ref{main++} it suffices to note that the strong convergence
of the sequence $\{\Phi_n\}$ to the operation $\Phi_0$ implies convergence of the sequences  $\{\Phi_n(\rho_n)\}$ and $\{\Phi_n(\sigma_n)\}$
to the operators $\Phi_0(\rho_0)$ and $\Phi_0(\sigma_0)$ correspondingly (due to the uniform boundedness of the operators norms of all the maps $\Phi_n$).\smallskip

\emph{Proof of Theorem \ref{main++}.} We will assume that  $D(\rho_n\|\shs\sigma_n)$ is finite for all $\,n\geq0$.\smallskip

If $\sigma_0=0$ then (\ref{in+}) can be valid only if $\rho_0=0$ and the limit in (\ref{in+}) is equal to zero. In this case
the validity of (\ref{out+}) follows from the monotonicity property (\ref{m-prop}) and the lower semicontinuity of the relative entropy.
So, we may consider in what follows that $\sigma_0\neq0$.\smallskip

At the first step we will assume that
$\{\Phi_n\}$ is a sequence of quantum operations from $\T(\H_A)$ to $\T(\H_B)$ strongly converging to a quantum operation $\Phi_0$, i.e.
we begin with proving the claim of Corollary \ref{main++c}.

By Lemma \ref{op-ch} in Section 2.3 there is a sequence $\{\widetilde{\Phi}_n\}$ of quantum channels  from $\T(\H_A)$ to $\T(\H_B\oplus\H_C)$ strongly converging to a quantum channel $\widetilde{\Phi}_0$ such that
$$
\Phi_n(\rho)=P_B\widetilde{\Phi}_n(\rho)P_B\quad \forall \rho\in\T(\H_A),\; \forall n\geq0,
$$
where $P_B$ is the projector onto the subspace $\H_B$ of $\mathcal{H}_{B}\oplus\mathcal{H}_{C}$.

Thus, by Lemma \ref{V-lemma} in Section 5.1, it suffices to prove (\ref{out-l-r-c}) assuming that $ \{\Phi_n\}$ is a sequence of quantum channels strongly converging to a quantum channel $\Phi_0$. In this case there exist a system $E$ and a sequence $\{V_n\}$  of isometries  from $\,\mathcal{H}_{A}$ into $\mathcal{H}_{BE}$ strongly converging to an isometry $V_0$ such that $\mathrm{\Phi}_n(\varrho)=\mathrm{Tr}_E V_n\varrho V^*_n$ for all $\,n\geq0$  \cite[Theorem 7]{CSR}. It is clear that the operators
$V_n\rho_nV_n^*$ and $V_n\sigma_nV_n^*$ tend, respectively, to the operators
$V_0\rho_0V_0^*$ and $V_0\sigma_0V_0^*$ as $n\to+\infty$. Since
all the operators  $V_n$ are isometries, it follows from  (\ref{in+}) that
\begin{equation*}
\lim_{n\to+\infty}D(V_n\rho_nV_n^*\|\shs V_n\sigma_nV_n^*)=D(V_0\rho_0V_0^*\|\shs V_0\sigma_0V_0^*)<+\infty.
\end{equation*}
By Theorem \ref{main} this relation implies (\ref{out-l-r-c}), since the partial trace over $\H_E$ is a channel. \smallskip

At the second step we will prove the claim of the theorem under the condition
\begin{equation}\label{f-s-assump}
  \supp \sigma_n \subseteq \supp \sigma_0\quad \forall n>0.
\end{equation}
By the finiteness of $D(\rho_n\|\shs\sigma_n)$ this condition  implies that
$\supp \rho_n\subseteq \supp \sigma_0$ for all $n\geq0$. So, in this case we may assume that
$\sigma_0$ is a faithful state, i.e. $\ker\sigma_0=\{0\}$.

To prove the claim of the theorem  it suffices to show that the assumption
\begin{equation}\label{n-assumpt}
\lim_{n\to+\infty}D(\varrho_n\|\shs \varsigma_n)=a\neq D(\varrho_0\|\shs \varsigma_0)
\end{equation}
leads to a contradiction.

The condition $\varsigma_n=\Phi_n(\sigma_n)$ implies, by the uniform boundedness
of the operator norms of all the maps $\Phi_n$, that
$$
\lim_{n\to+\infty}\Phi_n(\sigma_0)=\lim_{n\to+\infty}\varsigma_n+\lim_{n\to+\infty}\Phi_n(\sigma_0-\sigma_n)=\varsigma_0.
$$
Hence, the compactness criterion for families of quantum operations in the strong convergence topology presented in \cite[Corollary 2]{AQC}
allows us to conclude that the sequence $\{\Phi_n\}$ is relatively compact w.r.t. this topology.\footnote{There is a misprint in Corollary 2 in \cite{AQC}: the word "closed" should be added to the first line.} Thus, there exists
a subsequence $\{\Phi_{n_k}\}$ strongly  converging to a quantum operation $\Psi$.

By the first step of the proof the limit relation (\ref{in+}) implies that
\begin{equation*}
\lim_{k\to+\infty}D(\varrho_{n_k}\|\shs\varsigma_{n_k})=\lim_{k\to+\infty}D(\Phi_{n_k}(\rho_{n_k})\|\shs \Phi_{n_k}(\sigma_{n_k}))=D(\Psi(\rho_0)\|\shs \Psi_0(\sigma_0))<+\infty.
\end{equation*}
This contradicts to (\ref{n-assumpt}), since it is easy to see that
$$
\Psi(\rho_0)=\lim_{k\to+\infty}\Phi_{n_k}(\rho_{n_k})=\varrho_0\quad \textrm{and}\quad \Psi(\sigma_0)=\lim_{k\to+\infty}\Phi_{n_k}(\sigma_{n_k})=\varsigma_0.
$$

To complete the proof of the theorem we have to remove condition (\ref{f-s-assump}).\smallskip

Assume that the operator $\sigma_0$ has finite rank $r$. Without loss of generality
we may assume that $\rank\sigma_n\geq r$ for all $n$. Denote by $P^n_r$ the spectral
projector of the operator $\sigma_n$ corresponding to its maximal $r$ eigenvalues\footnote{see Remark \ref{spectral-pr} in Section 2.1 how to avoid the ambiguity of the definition of $P^n_r$ associated with multiple eigenvalues.} (taking the multiplicity into account).
Then $P^0_r$ is the projector onto the support of $\sigma_0$ that contains the support of $\rho_0$ (since otherwise $D(\rho_0 \|\shs\sigma_0)=+\infty$).
By using the arguments presented at the end of Section 4.2.1 in \cite{QC}
(based on Theorem VIII.23 in \cite{R&S} and the Mirsky inequality (\ref{Mirsky-ineq+})) it is easy to show that
\begin{equation}\label{P-r-lim}
   P^0_r=n\,\textup{-}\!\!\lim_{n\to+\infty} P^n_r,
\end{equation}
where the limit in the operator norm topology.

Thus, we may assume that $\|P^n_r-P^0_r\|<1$ for all $n$. For each $n\neq0$ consider the unitary
operator
$$
U_n=W(P^n_r,P^0_r)+V_n,
$$
where $W(P^n_r,P^0_r)$ is the partial isometry defined in Lemma \ref{s-lemma}A below and
$V_n$ is any partial isometry such that $\,V_n^*V_n=I_{A}-P^0_r$ and $\,V_nV_n^*=I_{A}-P^n_r$.
Consider the sequences of operators
$$
\tilde{\rho}_n=U^*_n\rho_nU_n\quad \textrm{and} \quad\tilde{\sigma}_n=U^*_n\sigma_nU_n.
$$
By using (\ref{P-r-lim}) and Lemma \ref{s-lemma}B below it is easy to show that
\begin{equation}\label{new-l-r}
\lim_{n\to+\infty}\tilde{\rho}_n=\rho_0\quad \textrm{and} \quad\lim_{n\to+\infty}\tilde{\sigma}_n=\sigma_0.
\end{equation}
Since (\ref{in+}) implies that
\begin{equation*}
\lim_{n\to+\infty}D(\tilde{\rho}_n\|\shs\tilde{\sigma}_n)=D(\rho_0\|\shs\sigma_0)<+\infty,
\end{equation*}
it follows from Lemma \ref{V-lemma} in Section 5.1 that
\begin{equation}\label{tmp-lr-1}
\lim_{n\to+\infty}D(P^0_r\tilde{\rho}_nP^0_r\|\shs P^0_r\tilde{\sigma}_n)=D(P^0_r\rho_0P^0_r\|\shs P^0_r\sigma_0)=D(\rho_0\|\shs\sigma_0)<+\infty
\end{equation}
and
\begin{equation}\label{tmp-lr-2}
\lim_{n\to+\infty}D(\bar{P}^0_r\tilde{\rho}_n\bar{P}^0_r\|\shs \bar{P}^0_r\tilde{\sigma}_n)=D(\bar{P}^0_r\rho_0\bar{P}^0_r\|\shs \bar{P}^0_r\sigma_0)=D(0\|\shs0)=0,
\end{equation}
where $\bar{P}^0_r=I_A-P^0_r$ and we have used that $P^0_r\tilde{\sigma}_n=\tilde{\sigma}_nP^0_r$ by the construction.

Note that
\begin{equation}\label{var-eq-1}
\varrho_n=\widetilde{\Phi}_n(\tilde{\rho}_n)\quad \textrm{and}\quad \varsigma_n=\widetilde{\Phi}_n(\tilde{\sigma}_n)\quad \forall n\neq0,
\end{equation}
where $\widetilde{\Phi}_n$ is the quantum operation $\Phi_n(U_n(\cdot)U_n^*)$. Consider the sequences
of operators
$$
\varrho'_n=\widetilde{\Phi}_n(P^0_r\tilde{\rho}_n P^0_r)\quad \textrm{and}\quad  \varsigma'_n=\widetilde{\Phi}_n(P^0_r\tilde{\sigma}_n).
$$
Since $P^0_r\sigma_0=\sigma_0 P^0_r=\sigma_0$ and $P^0_r\rho_0P_r^0=\rho_0$, it follows
from (\ref{new-l-r}) and (\ref{var-eq-1}) that the sequences $\{\varrho'_n\}$  and $\{\varsigma'_n\}$ converge, respectively,
to the operators  $\varrho_0$  and $\varsigma_0$.

Thus, since condition (\ref{f-s-assump}) with $\sigma_n=P^0_r\tilde{\sigma}_n$ holds, the limit relation (\ref{tmp-lr-1})  implies, by the second  step of the proof, that
\begin{equation}\label{tmp-lr-1+}
\lim_{n\to+\infty}D(\varrho'_n \|\shs\varsigma'_n)=\lim_{n\to+\infty}D(\widetilde{\Phi}_n(P^0_r\tilde{\rho}_nP^0_r)\|\shs \widetilde{\Phi}_n(P^0_r\tilde{\sigma}_n))=D(\varrho_0\|\shs\varsigma_0)<+\infty.
\end{equation}
The limit relation (\ref{tmp-lr-2}) and the monotonicity property (\ref{m-prop}) show that
\begin{equation}\label{tmp-lr-2+}
\lim_{n\to+\infty}D(\widetilde{\Phi}_n(\bar{P}^0_r\tilde{\rho}_n\bar{P}^0_r)\|\shs \widetilde{\Phi}_n(\bar{P}^0_r\tilde{\sigma}_n))=0.
\end{equation}
By Corollary 3 in \cite{DTL} it follows from (\ref{tmp-lr-1+}) and (\ref{tmp-lr-2+}) that
\begin{equation*}
\lim_{n\to+\infty}D(\widetilde{\Phi}_n(P^0_r\tilde{\rho}_nP^0_r+\bar{P}^0_r\tilde{\rho}_n\bar{P}^0_r)\|\shs \widetilde{\Phi}_n(\tilde{\sigma}_n))=D(\varrho_0\|\shs\varsigma_0).
\end{equation*}
Since  $\frac{1}{2}\tilde{\rho}_n\leq P^0_r\tilde{\rho}_nP^0_r+\bar{P}^0_r\tilde{\rho}_n\bar{P}^0_r$ (see the end of the proof of Theorem \ref{main}),
the last limit relation implies,  by Proposition 2 in \cite{DTL} that
\begin{equation*}
\lim_{n\to+\infty}D\!\left(\textstyle\frac{1}{2}\varrho_n\|\shs \varsigma_n\right)=\lim_{n\to+\infty}D\!\left(\widetilde{\Phi}_n(\textstyle\frac{1}{2}\tilde{\rho}_n)\|\shs \widetilde{\Phi}_n(\tilde{\sigma}_n)\right)=D\!\left(\textstyle\frac{1}{2}\varrho_0\|\shs\varsigma_0\right).
\end{equation*}
This shows, due to (\ref{D-mul}) and (\ref{D-c-id}), the validity of (\ref{out+}).\smallskip

Assume that the operator $\sigma_0$ has infinite rank. Consider the sequences of operators
$$
\tilde{\rho}_n \doteq W^*_n\rho_nW_n\quad \textrm{and} \quad\tilde{\sigma}_n \doteq W^*_n\sigma_nW_n
$$
defined by means of the sequence $\{W_n\}$ of partial isometries with the properties stated in Lemma
\ref{r-lemma} below. Since $\supp\rho_n\subseteq\supp\sigma_n$  due to the assumed finiteness of $D(\rho_n\|\shs\sigma_n)$, we have
$\supp\tilde{\rho}_n\subseteq\supp\tilde{\sigma}_n\subseteq\H_0\doteq\supp\sigma_0\,$ and $D(\tilde{\rho}_n\|\shs\tilde{\sigma}_n)=D(\rho_n\|\shs\sigma_n)$ for all $n$ by the construction.

The assumption (\ref{in+}) shows that
$$
\sup_{n\geq n_0} D(\tilde{\rho}_n\|\shs\tilde{\sigma}_n)=\sup_{n\geq n_0} D(\rho_n\|\shs\sigma_n)<+\infty
$$
for some $n_0\in\mathbb{N}$. Since the set $\{\tilde{\sigma}_n\}\cup\{\sigma_0\}$ is compact due to the second  limit relation in (\ref{r-lemma+}), this implies, by Lemma \ref{rho-comp} below, that
the sequence $\{\tilde{\rho}_n\}$ is relatively compact. Assume that $\rho_*$ is a partial limit of $\{\tilde{\rho}_n\}$, i.e.
there is a subsequence $\{\tilde{\rho}_{n_k}\}$ converging to $\rho_*$.   Then the
first limit relation in (\ref{r-lemma+}) shows that
$$
\rho_0=\lim_{k\to+\infty}\rho_{n_k}=\lim_{k\to+\infty} W_{n_k}\tilde{\rho}_{n_k}W_{n_k}^*=R_0\rho_*R_0=\rho_*,
$$
where $R_0$ is the projector onto $\H_0$ and the last equality holds as the support of $\rho_*$  lies within $\H_0$
(since the supports of all the operators $\tilde{\rho}_n$ are contained in $\H_0$). Thus, all
the partial limits of the relatively compact sequence $\{\tilde{\rho}_n\}$ coincide with $\rho_0$.

Thus, the sequences $\{\tilde{\rho}_n\}$ and $\{\tilde{\sigma}_n\}$ converge, respectively, to the operators $\rho_0$ and
$\sigma_0$ and
\begin{equation*}
\lim_{n\to+\infty}D(\tilde{\rho}_n\|\shs\tilde{\sigma}_n)=D(\rho_0\|\shs\sigma_0)<+\infty.
\end{equation*}
Note that $\varrho_n=\widetilde{\Phi}_n(\tilde{\rho}_n)$ and $\varsigma_n=\widetilde{\Phi}_n(\tilde{\sigma}_n)$
for each $n\neq0$, where $\widetilde{\Phi}_n$ is the quantum operation $\Phi_n(W_n(\cdot)W_n^*)$. Moreover,
the sequence $\{\tilde{\sigma}_n\}$ satisfies condition (\ref{f-s-assump}). Hence, the second step of the proof shows the validity of (\ref{out+}).\smallskip

\begin{lemma}\label{s-lemma} \emph{For finite rank
projectors $P$ and $Q$ on a space $\H$ denote by $W(P,Q)$ the partial isometry from the polar decomposition
of the operator $PQ$, i.e. $W(P,Q)$  is the unique operator such that $\,W(P,Q)W(P,Q)^*$ and $\,W(P,Q)^*W(P,Q)$
are the projectors onto the range of $PQ$ and $\sqrt{QPQ}$ correspondingly and $\,PQ=W(P,Q)\sqrt{QPQ}$.}\smallskip

A) \emph{If $\|P-Q\|<1$ then  $\,W(P,Q)W(P,Q)^*=P$ and $\,W(P,Q)^*W(P,Q)=Q$.}\smallskip

B) \emph{If $\{P_n\}$ is a sequence of projectors converging to a projector $P_0$ in the operator norm then
the sequence  $\{W(P_n,P_0)\}$ converges to the projector $P_0$ in the operator norm.}
\end{lemma}\smallskip

\emph{Proof.} A) It is clear that $\mathrm{R} (PQ)\subseteq\mathrm{R} (P)$, where $\mathrm{R}(X)$ denotes the range of $X$. Assume that $\varphi$ is a unit
vector in $\mathrm{R}(P)\ominus\mathrm{R} (PQ)$. Then for any vector $\psi$ in $\H$ we have
$\langle\varphi |Q\psi\rangle=\langle P\varphi |Q\psi\rangle=\langle\varphi |PQ\psi\rangle=0$,
i.e. $\varphi$ lies in $[\mathrm{R}(Q)]^\perp$. This contradicts to the condition $\|P-Q\|<1$.

Thus, $\mathrm{R} (PQ)=\mathrm{R} (P)$. Note that  $\mathrm{R} (\sqrt{QPQ})=\mathrm{R}(QPQ)\subseteq\mathrm{R} (Q)$.  Assume that $\varphi$ is a unit
vector in $\mathrm{R} (Q)\ominus\mathrm{R} (QPQ)$. Then for any vector $\psi$ in $\H$ we have
$\langle\varphi |PQ\psi\rangle=\langle Q\varphi |PQ\psi\rangle=\langle\varphi |QPQ\psi\rangle=0$,
i.e. $\varphi$ lies in $[\mathrm{R}(PQ)]^\perp=[\mathrm{R}(P)]^\perp$. This contradicts to the condition $\|P-Q\|<1$.
Hence $\mathrm{R}(\sqrt{QPQ})=\mathrm{R}(Q)$.\smallskip

B)  Since the sequences $\{P_nP_0\}$ and $\{\sqrt{P_0P_nP_0}\}$ converges to the projector $P_0$ in the operator norm \cite{R&S}, it is easy to show that the sequence
$\{W(P_n,P_0)\}$ converges to the projector $P_0$ in the operator norm as well. $\Box$

\smallskip

\begin{lemma}\label{r-lemma}  \emph{Let $\{\sigma_n\}$ be a sequence in $\T_+(\H)$ converging to an infinite rank operator $\sigma_0$.
Then there is a sequence $\{W_n\}$ of partial isometries such that $W_n^*W_n=R_0$, $W_nW_n^*\geq R_n$,
\begin{equation}\label{r-lemma+}
 s\shs\textup{-}\lim_{n\rightarrow\infty} W_n=R_0\quad \textit{and} \quad \lim_{n\rightarrow\infty} W^*_n\sigma_nW_n=\sigma_0,
\end{equation}
where $R_0$ and $R_n$ are the projectors onto the supports of the operators $\sigma_0$ and $\sigma_n$ and $s\shs\textup{-}\lim$ denotes the limit in the strong operator topology.}
\end{lemma}\smallskip

\emph{Proof.} We may assume w.l.o.g. that the sequence $\{\sigma_n\}$ consists of states and converges to an infinite rank state $\sigma_0$. \smallskip

Denote by $\{\lambda^{n}_i\}_{i=1}^{+\infty}$ the sequence
of eigenvalues of $\sigma_n$ in the non-increasing order (taking the multiplicity into account). By the Mirsky inequality (\ref{Mirsky-ineq+})
we have
\begin{equation}\label{Mirsky-ineq}
  \sum_{i=1}^{+\infty}|\lambda^{n}_i-\lambda^{0}_i|\leq \|\sigma_n-\sigma_0\|_1\quad \forall n.
\end{equation}

For each natural $j\leq\rank\sigma_n$ and $n\geq0$ let $P^n_j$ be the spectral
projector of the state $\sigma_n$ corresponding to the eigenvalues $\lambda^{n}_1,...,\lambda^{n}_j$.\footnote{see Remark \ref{spectral-pr} in Section 2.1 how to avoid the ambiguity of the definition of $P^n_j$ associated with multiple eigenvalues.}  If $\rank\sigma_n< j$ then $P^n_j=R_n$.

Let $\{\mu_k\}_{k=1}^{+\infty}$ be a sequence of \emph{different} eigenvalues of $\sigma_0$ taken in the non-increasing order.
Denote by $m_k$ the multiplicity of $\mu_k$. For each $n\geq0$ introduce the projectors
$$
Q^n_1=P^n_{m_1},\quad Q^n_k=P^n_{m_1+m_2+...+m_k}-P^n_{m_1+m_2+...+m_{k-1}},\; k=2,3,...
$$
Since
\begin{equation}\label{inc-rank}
  \lim_{p\to+\infty}\min_{n\geq p}\rank\sigma_n=\rank\sigma_0=+\infty,
\end{equation}
by using Theorem VIII.23 in \cite{R&S} and  inequality (\ref{Mirsky-ineq}) it is easy to show that
\begin{equation}\label{Q-conv}
  n\,\textup{-}\!\!\lim_{n\rightarrow+\infty} Q_k^n=Q_k^0\qquad\forall k\in\mathbb{N},
\end{equation}
where the limit in the operator norm topology.

For any two projectors $P$ and $Q$ of the same finite rank we will denote by
$W(P,Q)$ the partial isometry from $Q(\H)$ onto $P(\H)$ introduced in Lemma \ref{s-lemma}A
if $\|P-Q\|<1$ and any partial isometry from $Q(\H)$ onto $P(\H)$ otherwise. \smallskip

If for given $n$ the state $\sigma_n$ has infinite rank
then both projectors $Q_k^n$ and $Q_k^0$ have the same rank $m_k$ for all $k$ and we
define the operator $W_n$ as follows
$$
W_n=\sum_{k=1}^{+\infty}W(Q_k^n,Q_k^0),
$$
where $W(Q_k^n,Q_k^0)$ is the partial isometry from $Q_k^0(\H)$ onto $Q_k^n(\H)$ defined above.
It is easy to see that $W_n$ is a partial isometry such that $W_n^*W_n=R_0$ and $W_nW_n^*=R_n$.

If for given $n$ the state $\sigma_n$ has finite rank $r_n>0$
then we denote by $k_n$ the maximal number in the set $\{k\in\mathbb{N}\,|\,\rank Q_k^n=m_k\}$ provided that this set is not empty and assume that $k_n=0$ otherwise. We
define the operator $W_n$ as follows
$$
W_n=\sum_{k=1}^{k_n}W(Q_k^n,Q_k^0)+W(Q'_n,Q_{k_{n}+1}^0)+V_{n},
$$
where
\begin{description}
  \item[-] $Q'_n$ is any $m_{k_{n}+1}$-rank projector such that $Q_{k_{n}+1}^n\leq Q'_n\leq I_{\H}-\sum_{k=1}^{k_n}Q_k^n$ if
$r_n>m_1+...+m_{k_n}$,
  \item[-] $Q'_n$ is any $m_{k_{n}+1}$-rank projector s.t. $Q'_{n}\leq I_{\H}-\sum_{k=1}^{k_n}Q_k^n\,$ if
$\,r_n=m_1+...+m_{k_n}$,
  \item[-] $V_n$ is any partial isometry such that $V_n^*V_n=R_0-\sum_{k=1}^{k_{n}+1}Q_k^0$ and $V_nV_n^*=I_{\H}-\sum_{k=1}^{k_n}Q_k^n-Q'_n$
(such operator exists, since both projectors $R_0-\sum_{k=1}^{k_{n}+1}Q_k^0$ and $I_{\H}-\sum_{k=1}^{k_n}Q_k^n-Q'_n$ have infinite rank),
\end{description}
and it is assumed that $\,\sum_{k=1}^{k_n}(\cdot)=0\,$ and  $\,m_1+m_2+...+m_{k_n}=0\,$  if $\,k_n=0$.\smallskip

It is easy to see that $W_n$ is a partial isometry such that $W_n^*W_n=R_0$ and  $W_nW_n^*=I_{\H}$.

Since $W_nR_0=W_n$ for all $n$,  to prove the first limit relation in (\ref{r-lemma+}) it suffices
to show that $\lim_{n\rightarrow+\infty} W_n|\varphi\rangle=|\varphi\rangle$
for any vector $\varphi$ such that $|\varphi\rangle=\sum_{k=1}^{k_\varphi}Q_k^0|\varphi\rangle$ for some
finite $k_\varphi$ depending on $\varphi$. This can be done by noting that
\begin{equation}\label{W-conv}
W_n|\varphi\rangle=\sum_{k=1}^{k_\varphi}[W(Q_k^n,Q_k^0)]|\varphi\rangle
\end{equation}
for all $n$ large enough due to (\ref{inc-rank}).  It follows from (\ref{Q-conv}) and Lemma \ref{s-lemma}B
that the r.h.s. of (\ref{W-conv}) tends to $|\varphi\rangle$ as $n\to+\infty$.

To prove the second limit relation in (\ref{r-lemma+}) it suffices to note that
$\|W^*_n\sigma_nW_n-\sigma_0\|_1$ coincides with the l.h.s. of (\ref{Mirsky-ineq}) by
the construction. $\Box$ \medskip

\begin{lemma}\label{rho-comp}  \emph{If $\,\C$ is a compact subset of $\,\T_+(\H)$ then
the set $\T_{b,c}$ of all operators $\rho$ in $\T_+(\H)$ such that
$\Tr \rho\leq b$ and $\,\inf_{\sigma\in\C} D(\rho\shs\|\sigma)\leq c$
is compact w.r.t. the trace norm for any positive numbers $b$ and $c$.}
\end{lemma}\smallskip

\emph{Proof.} The lower semicontinuity of the relative entropy and the compactness of $\C$ show that
for any operator $\rho$ in $\T_{b,c}$ there is a state $\sigma_{\rho}$ in $\C$ such that
$D(\rho\shs\|\sigma_{\rho})\leq c$. So, by repeatedly using the lower semicontinuity of the relative entropy and the compactness of $\C$
it is easy to show the closedness of $\T_{b,c}$.

To prove the relative compactness of $\T_{b,c}$ note that for any (orthogonal) projector $P$
the basic properties the relative entropy (cf.\cite{L-2,L-REM}) imply that
\begin{equation}\label{D-ineq}
D(\Tr P\rho P\shs\|\Tr P\sigma_{\rho} P)\leq D(P\rho P\shs\|P\sigma_{\rho} P)\leq D(\rho\shs\|\sigma_{\rho})\leq c
\end{equation}
for any operator $\rho$ in $\T_{b,c}$, where
$$
D(\Tr P\rho P\shs\|\Tr P\sigma_{\rho} P)\doteq\Tr P\rho \ln (\Tr P\rho/\Tr P\sigma_{\rho})+\Tr P\sigma_{\rho}-\Tr P\rho
$$
is the relative entropy  between the nonnegative numbers $\Tr P\rho$ and $\Tr P\sigma_{\rho}$ (that can be treated as operators on one-dimensional Hilbert space).

By the compactness criterion for bounded subsets of $\T_+(\H)$ (\cite[the Appendix]{AQC})
the compactness of $\C$ shows that for any $\varepsilon>0$ there is a finite rank projector $P_\varepsilon$
such that $\Tr (I_{\H}-P_\varepsilon)\sigma<\varepsilon$ for all $\sigma$ in $\C$. So, it follows from inequality (\ref{D-ineq})
that
$$
\Tr (I_{\H}-P_\varepsilon)\rho\leq \frac{b+c+1/e}{-\ln\Tr (I_{\H}-P_\varepsilon)\sigma_{\rho}}\leq \frac{b+c+1/e}{-\ln \varepsilon}
$$
for any operator $\rho$ in $\T_{b,c}$. Thus, since the r.h.s. of the last inequality can be made arbitrarily small
by choosing sufficiently small $\varepsilon$, the compactness criterion mention before implies
the relative compactness of $\T_{b,c}$. $\Box$\bigskip

\section{Concluding remarks}

It is well known that the use of quantum relative entropy often helps to overcome the analytical difficulties that arise when studying infinite-dimensional quantum systems and channels. For example,  replacing the expression
$$
I(A\!:\!B)_{\omega}=S(\omega_A)+S(\omega_B)-S(\omega)
$$
for the quantum mutual information of a state $\omega$ of a bipartite system $AB$
by the expression
$$
I(A\!:\!B)_{\omega}=D(\omega\|\shs\omega_A\otimes\omega_B)
$$
allows us to correctly define this quantity for all states in $\S(\H_{AB})$ in the case of infinite-dimensional systems $A$ and $B$ and to establish its basic properties \cite{L-mi}.
The same situation occurs with other important characteristics of quantum systems and channels such as
the Holevo quantity, the mutual and coherent information of a channel, etc.

The preservation of local continuity of the quantum relative entropy under completely positive linear maps established in this article can be considered as another manifestation of the above observation. Indeed, it suffices to note that local continuity of von Neumann entropy \emph{is not preserved} by completely positive linear maps (in particular, by quantum channels), i.e.
the existence of
\begin{equation}\label{S-in}
 \lim_{n\to+\infty}S(\rho_n)=S(\rho_0)<+\infty
\end{equation}
for a sequence $\{\rho_n\}\subset\S(\H_A)$ converging to a state $\rho_0$ does not imply
\begin{equation}\label{S-out}
 \lim_{n\to+\infty}S(\Phi(\rho_n))=S(\Phi(\rho_0))<+\infty
\end{equation}
for a quantum channel $\Phi:A\to B$ of general type. The implication (\ref{S-in})$\Rightarrow$(\ref{S-out})
holds if and only if the function $\varrho\mapsto S(\Phi(\varrho))$ is bounded on the set of pure input states \cite[Theorem 1]{PFE}. It is clear that
the last condition is very restrictive and does not hold for most quantum channels used in applications.

\bigskip

I am grateful to A.S.Holevo and to the participants of his seminar
"Quantum probability, statistic, information" (the Steklov
Mathematical Institute) for useful discussion.
I am also grateful to G.G.Amosov for the question motivated this research (it was mentioned in the Introduction).
Special thanks to A.Winter for the useful comment concerning Mirsky's inequality.
\smallskip



\bigskip

\begin{thebibliography}{99}

\bibitem{IRE-1} B.Schumacher, M.D.Westmoreland, "Relative entropy in quantum information theory",
arXiv:quant-ph/0004045.

\bibitem{IRE-2} V. Vedral,  "The Role of Relative Entropy in Quantum Information Theory",
Rev. Mod. Phys. \textbf{74}, 197;  arXiv:quant-ph/0102094.

\bibitem{W} A.Wehrl, "General properties of entropy", Rev. Mod. Phys. \textbf{50},
221-250 (1978).

\bibitem{O&P} M.Ohya, D.Petz, Quantum Entropy and Its Use, Theoretical and Mathematical Physics (Springer Berlin Heidelberg,
2004).

\bibitem{H-SCI} A.S.Holevo "Quantum systems, channels, information.
A mathematical introduction", Berlin, DeGruyter, 2012.

\bibitem{Wilde} M.M.Wilde, "Quantum Information Theory".  Cambridge University Press, 2013.


\bibitem{L-2} G.Lindblad  "Expectation and Entropy Inequalities for Finite
Quantum Systems", Comm. Math. Phys. 1974. \textbf{39}:2, 111-119.


\bibitem{Nau} J.Naudts, "Continuity of a class of entropies and relative entropies", Rev. Math. Phys. \textbf{16}:6, 809-822 (2004); arXiv:math-ph/0208038.

\bibitem{DTL} M.E.Shirokov, "Convergence conditions for the quantum relative entropy and other applications of the deneralized quantum Dini lemma", Lobachevskii J. Math. \textbf{43}, 1755-1777 (2022); arXiv:2205.09108.

\bibitem{BSimon} B.Simon, "Operator Theory: A Comprehensive Course in Analysis, Part IV",  American Mathematical Society, 2015.

\bibitem{Mirsky} L. Mirsky, "Symmetric gauge functions and unitarily invariant norms", Quart. J. Math.Oxford \textbf{2}(11), 50-59 (1960).

\bibitem{N&Ch} M.A.Nielsen, I.L.Chuang, "Quantum Computation and Quantum
Information", Cambridge University Press, 2000.

\bibitem{Rudin} W.R.Rudin, "Principles of Mathematical Analysis", McGraw-Hill, 1976.

\bibitem{Ume} H.Umegaki, "Conditional expectation in an operator algebra, IV (entropy and information)", Kodai Math.Sem.Rep., \textbf{14}, 59-85 (1962).

\bibitem{B&R} O.Bratteli, D.W.Robinson, "Operators algebras and quantum statistical mechanics", vol.I, Springer Verlag, New York-Heidelberg-Berlin, 1979.

\bibitem{R&S} M.Reed, B.Simon, "Methods of Modern Mathematical Physics. Vol I. Functional Analysis", Academic Press Inc., 1980.

\bibitem{AQC} A.S.Holevo, M.E.Shirokov "On approximation of
infinite dimensional quantum channels", Problems of Information
Transmission. 2008. \textbf{44}:2,  3-22; arXiv:0711.2245.



\bibitem{D-A} G.F.Dell'Antonio, "On the limits of sequences of normal states", Commun.
Pure Appl. Math. \textbf{20}, 413-430 (1967).

\bibitem{SSC} M.E.Shirokov, "Strong* convergence of quantum channels", Quantum Inf. Process., \textbf{20}, 145, 16 pp. (2021); arXiv:1802.05632.

\bibitem{W-EBN} A.Winter, "Energy-Constrained Diamond Norm with Applications to the Uniform Continuity of Continuous Variable Channel Capacities",  arXiv:1712.10267.

\bibitem{CSR} M.E.Shirokov, "Strong convergence of quantum channels: continuity of the Stinespring dilation and discontinuity of the unitary dilation", J. Math. Phys. 2020. \textbf{61}, 082204; arXiv:1712.03219.

\bibitem{EC} M.E.Shirokov, "Entropy characteristics of subsets of states I",  Izv. Math., \textbf{70}:6, 1265-1292  (2006); arXiv: quant-ph/0510073.

\bibitem{L-REM} G.Lindblad  "Completely Positive Maps
and Entropy Inequalities", Comm. Math. Phys. \textbf{40}:2,147-151 (1975).

\bibitem{QC} M.E.Shirokov, "Continuity of characteristics of composite quantum systems: a review", arXiv:2201.11477.

\bibitem{L-mi} G.Lindblad,  "Entropy, information and quantum measurements", Comm. Math. Phys. \textbf{33}, 305-322 (1973).


\bibitem{PFE} A.V.Bulinski, M.E.Shirokov, "On quantum channels and operations preserving finiteness of the von Neumann entropy", Lobachevskii J. Math., \textbf{41}:12, 2383-2396 (2020); arXiv:2004.03582.




\end{thebibliography}
\end{document}